\documentclass[10pt,a4paper]{article}
\usepackage[margin=17mm,columnsep=6mm]{geometry}
\usepackage{authblk}
\usepackage[numbers,sort&compress]{natbib}
\usepackage{microtype}
\usepackage{amsthm} 
\usepackage{titlesec}
\usepackage[font=scriptsize,labelfont=bf]{caption}
\usepackage{booktabs}
\usepackage{subfigure}
\usepackage{enumitem}
\usepackage{rotating}
\usepackage{ragged2e}
\usepackage{graphicx}
\graphicspath{{./figures/}}
\usepackage{float}
\usepackage{placeins}
\usepackage{graphicx}
\usepackage[table,dvipsnames]{xcolor}
\usepackage{amsmath,amssymb,amsfonts}
\usepackage[switch,mathlines]{lineno}

\newif\ifimvlinenumbers
\imvlinenumbersfalse

\usepackage{multirow}
\usepackage{booktabs}
\usepackage{algorithm}
\usepackage{algorithmic}
\usepackage{listings}
\usepackage{xcolor}
\usepackage[T1]{fontenc}
\usepackage{newtxtext,newtxmath}
\usepackage[varqu]{zi4}

\usepackage[framemethod=TikZ]{mdframed}
\definecolor{imvcodered}{RGB}{153,70,54}
\newcommand{\imvcodefont}{\fontsize{7}{8}\selectfont}
\newcounter{imvcodeblock}
\newenvironment{imvcodeblock}[2]{%
    \refstepcounter{imvcodeblock}\label{#2}%
    \begin{mdframed}[
        frametitle={\colorbox{white}{\space Box \textcolor{imvcodered}{\theimvcodeblock}: #1\space}},
        innertopmargin=0pt,
        frametitleaboveskip=-\ht\strutbox,
        frametitlealignment=\center,
        linewidth=1pt,
        roundcorner=10pt
    ]%
}{%
    \end{mdframed}%
}

\definecolor{codeblue}{rgb}{0.1,0.1,0.6}
\definecolor{codegreen}{rgb}{0,0.6,0}
\definecolor{codegray}{rgb}{0.5,0.5,0.5}
\definecolor{codepurple}{rgb}{0.58,0,0.82}

\lstdefinestyle{mystyle}{
    backgroundcolor=\color{white},   
    commentstyle=\color{codegreen},
    keywordstyle=\color{codeblue},
    numberstyle=\tiny\color{codegray},
    stringstyle=\color{codepurple},
    basicstyle=\ttfamily\footnotesize,
    breakatwhitespace=false,         
    breaklines=true,                 
    captionpos=b,                    
    keepspaces=true,                 
    numbers=left,                    
    numbersep=5pt,                  
    showspaces=false,                
    showstringspaces=false,
    showtabs=false,                  
    tabsize=2
}

\definecolor{tangokw}{RGB}{32,74,135}
\definecolor{tangostr}{RGB}{78,154,6}
\definecolor{tangocom}{RGB}{143,89,2}
\lstdefinestyle{imvpython}{
    language=Python,
    basicstyle=\imvcodefont\ttfamily,
    keywordstyle=\bfseries\color{tangokw},
    stringstyle=\color{tangostr},
    commentstyle=\color{tangocom},
    numbers=none,
    breaklines=true,
    breakatwhitespace=false,
    columns=fullflexible,
    keepspaces=true,
    showstringspaces=false,
    upquote=true,
    aboveskip=0pt,
    belowskip=0pt,
    xleftmargin=0pt,
    frame=none
}

\usepackage[colorlinks=true,linkcolor=MidnightBlue,
    urlcolor=MidnightBlue,citecolor=MidnightBlue]{hyperref}
\hypersetup{
    pdftitle={Mastering Model Fit: Applications of the InterModel Vigorish for Interpretable Machine Learning},
    pdfauthor={Jiani Yan, Pengyun Wang, Klint Kanopka, Ben Domingue, Charles Rahal}
}
\titleformat*{\section}{\large\bfseries}
\titleformat*{\subsection}{\normalsize\bfseries}
\titlespacing*{\section}{0pt}{12pt plus 2pt minus 2pt}{6pt plus 1pt minus 1pt}
\titlespacing*{\subsection}{0pt}{8pt plus 2pt minus 1pt}{4pt plus 1pt minus 1pt}
\titleformat{\subsubsection}[runin]{\normalfont\itshape}
    {\thesubsubsection}{0.6em}{}[.]
\titlespacing*{\subsubsection}{0pt}{0.5\baselineskip}{0.6em}

\title{\LARGE\bfseries Mastering Model Fit: Applications of the InterModel Vigorish for Interpretable Machine Learning}
\author[1,2]{Jiani Yan\thanks{ Denotes joint lead authorship. \textsuperscript{\ensuremath{\dagger}} Denotes corresponding authorship. Corresponding authors: Jiani Yan, Leverhulme Centre for Demographic Science, University of Oxford, 42--43 Park End Street, Oxford OX1 1JD, United Kingdom; email: \href{mailto:jiani.yan@demography.ox.ac.uk}{jiani.yan@demography.ox.ac.uk}. Charles Rahal, Leverhulme Centre for Demographic Science, University of Oxford, 42--43 Park End Street, Oxford OX1 1JD, United Kingdom; email: \href{mailto:charles.rahal@demography.ox.ac.uk}{charles.rahal@demography.ox.ac.uk}.~We~gratefully acknowledge comments from participants at the 6th International Conference on Social Computing, and funding from the Leverhulme Trust (Grant RC-2018-003) for the Leverhulme Centre for Demographic Science, Nuffield College, the University of Chicago, the Jacobs Foundation, and the Economics and Social Research Council.}\textsuperscript{\ensuremath{\dagger},}}
\author[3]{Pengyun Wang\textsuperscript{\ensuremath{*},}}
\author[4]{Klint Kanopka}
\author[5]{Ben Domingue}
\author[1,2]{Charles Rahal\textsuperscript{\ensuremath{\dagger,}}}
\affil[1]{Leverhulme Centre for Demographic Science, University of Oxford}
\affil[2]{Nuffield College, University of Oxford}
\affil[3]{Data Science Institute, University of Chicago}
\affil[4]{NYU Steinhardt, New York University}
\affil[5]{Stanford Graduate School of Education, Stanford University}
\date{}

\usepackage{etoolbox}
\patchcmd{\abstract}{\begin{center}}{\begin{center}\vspace*{-15pt}}{}
  {\PackageError{IMV}{Could not adjust abstract spacing}{Check the article abstract environment.}}
\makeatletter
\patchcmd{\AB@maketitle}{\@thanks}{\global\let\imv@titlethanks\@thanks}{}
  {\PackageError{IMV}{Could not defer title notes}{Check the authblk title macro.}}
\AddToHook{begindocument/before}{\@LN{0}{0}}
\newcommand{\imvinternallinenumbers}{%
  \let\imv@linecounter\c@linenumber
  \internallinenumbers
  \let\c@linenumber\imv@linecounter
  \ifx\c@linenumber\c@pagewiselinenumber
    \def\theLineNumber{\number\numexpr\c@linenumber+1\relax}%
  \else
    \def\theLineNumber{\thelinenumber}%
  \fi
  \let\LineNumber\theLineNumber
  \let\makeLineNumber\makeLineNumberLeft}
\newcommand{\imvprinttitlethanks}{%
  \begingroup
  \renewcommand{\thefootnote}{\fnsymbol{footnote}}%
  \let\imv@makefntext\@makefntext
  \long\def\@makefntext##1{%
    \ifimvlinenumbers\imvinternallinenumbers\fi
    \imv@makefntext{##1}\par}%
  \imv@titlethanks
  \endgroup}
\makeatother

\begin{document}
\ifimvlinenumbers\leftlinenumbers\fi
\twocolumn[{
\begin{@twocolumnfalse}
\ifimvlinenumbers\imvinternallinenumbers\fi
\maketitle

\begin{abstract}
Accuracy alone gives little information about which features or model components improve probabilistic fit, which outcomes benefit, or how fit changes during training. We introduce the InterModel Vigorish (IMV) into the Machine Intelligence space in order to express differences in fit as an expected betting advantage over a reference model. SHAP-IMV attributes gains to features through Shapley-weighted subset comparisons, Multi-IMV evaluates one-vs-rest and pairwise class discrimination, and Ablate-IMV assesses architectural components and feature groups by ablation. We evaluate these applications over ten random seeds using tabular data, images, sensor sequences, and graphs. Feature rankings vary across estimators. High one-vs-rest accuracy can accompany negligible improvement over a null, while ablation can preserve or improve fit under a fixed training budget. On six graph-classification benchmarks, we time-profile IMV over 200 training epochs, reporting seed means and ranges. The open-source \texttt{imvpy} library and accompanying notebooks provide code to replicate our work and to use the IMV more widely.
\end{abstract}
\vspace{15pt}
\end{@twocolumnfalse}
}]
\thispagestyle{empty}
\ifimvlinenumbers\linenumbers\fi
\imvprinttitlethanks

\ifimvlinenumbers\switchlinenumbers\fi

\section{Introduction}
Two classifiers can make identical class predictions yet assign different probabilities to the observed outcomes. In applications of machine learning across disciplines~\cite{jordan2015machine,carleo2019machine,rahal2024rise,lin2024machine}, choosing between them requires a measure suited to the comparison and an appropriate evaluation of generalization~\cite{cawley2010over,yang2023test}. Measures such as F1~\cite{van1979information}, R${^2}$~\cite{R^2}, cross-entropy~\cite{rumelhart1986learning}, and AIC and BIC~\cite{AIC,BIC} answer different questions. Threshold-based scores summarize classification decisions, probabilistic scores assess the probabilities assigned to outcomes, and information criteria account for complexity given a specified likelihood. Each can be used to compare models, but a difference in scores need not have the same interpretation relative to a baseline~\cite{domingue2021intermodel}.

We adapt the InterModel Vigorish (IMV)~\cite{domingue2021intermodel,domingue2024intermodel} --- a model-agnostic measure for comparing binary probabilistic predictions --- to three tasks. SHAP-IMV uses Shapley values~\cite{shapley1953value,SHAP} to attribute gains in fit to features, conditional on the estimator, baseline, and feature set. Multi-IMV evaluates multiclass predictions through one-vs-all and one-vs-one comparisons~\cite{knerr1990single}, showing how fit differs across individual classes and class pairs. In Ablate-IMV, we remove components, refit the resulting models, and compare their probabilistic fit with that of the full model.

We examine logistic regression~\cite{hosmer2013applied}, XGBoost~\cite{chen2016xgboost}, LightGBM~\cite{ke2017lightgbm}, convolutional and bidirectional recurrent networks, multilayer perceptrons, and graph convolutional networks~\cite{kipf2016semi}. The experiments cover binary and multiclass tabular prediction, image parity, human activity recognition, and graph classification. Across these settings, IMV expresses gains in probabilistic fit relative to a reference model. Computing the score adds little cost once predictions are available; exact feature attribution is more demanding because it requires repeated model fitting.

\section{Related Work}
\label{sec::related work}
\subsection{Feature Importance and Local Explanations}
Model evaluation concerns both overall predictive performance and the contribution of individual predictors~\cite{sokolova2009systematic,huang2005using,powers2020evaluation}. Feature-importance methods address the latter either globally, or for individual predictions~\cite{SHAP}. Information Gain measures information gained at tree splits~\cite{Entropy,decisiontrees}, while Gini Importance measures reductions in impurity, commonly in random forests~\cite{Randomforests}. Mutual Information measures bivariate dependence~\cite{Entropy}. These quantities describe particular algorithms or relationships between variables; they do not directly measure gains in held-out probabilistic fit. For individual predictions, SHAP assigns contributions to features~\cite{SHAP,strumbelj2010efficient}, LIME approximates the fitted model with a local linear surrogate~\cite{LIME}, and Anchors constructs high-precision prediction rules~\cite{Anchors,hase2020evaluating}. The local scope of these explanations can leave global behavior unexamined. SAGE uses a loss-based value function to allocate predictive power through global Shapley attribution~\cite{covert2020sage}. 
Our SHAP-IMV also uses global Shapley attribution, but refits models for each
feature subset and defines the value as held-out IMV relative to a baseline,
so allocations are expressed in IMV units and sum to the full model's gain.

\subsection{Global Metrics for Machine Learning}
Brier score~\cite{brier1950verification}, accuracy~\cite{Accuracy}, precision, recall, F1~\cite{flach2015precision}, and ROC area~\cite{hanley1982meaning} summarize overall predictive performance~\cite{gordon2021disagreement}. IMV gives comparisons of probabilistic fit an explicit reference and a betting-scale interpretation~\cite{domingue2021intermodel}. Likelihood-ratio tests and information criteria provide other approaches to model comparison~\cite{AIC,BIC}. Whichever measure is used, claims about generalization depend on the evaluation design. Here we use IMV to attribute gains in probabilistic fit to features, to
evaluate multiclass predictions, and to assess the effects of model ablation.

\space
\section{Preliminaries}
\subsection{Problem Definition}
For binary classification~\cite{domingue2021intermodel}, let dataset $D$ contain $N$ observations with outcomes $Y_i\in\{0,1\}$ and $d$ features $\mathbf{X}_i=(X_{i1},\ldots,X_{id})$. Model $f$ assigns probability $p_i=f(\mathbf{X}_i)\in(0,1)$ to $Y_i=1$. The observed outcome contributes likelihood $L_i=p_i^{Y_i}(1-p_i)^{1-Y_i}$ and log-likelihood $\ell_i=Y_i\log p_i+(1-Y_i)\log(1-p_i)$. We summarize fit by the geometric mean likelihood:
\begin{equation}
 L(p; D) \equiv \left(\prod_{i=1}^{N} p_i^{Y_i} (1 - p_i)^{1 - Y_i}\right)^{\frac{1}{N}}.
\end{equation}

\begin{figure*}
    \centering
    \includegraphics[width=1\linewidth]{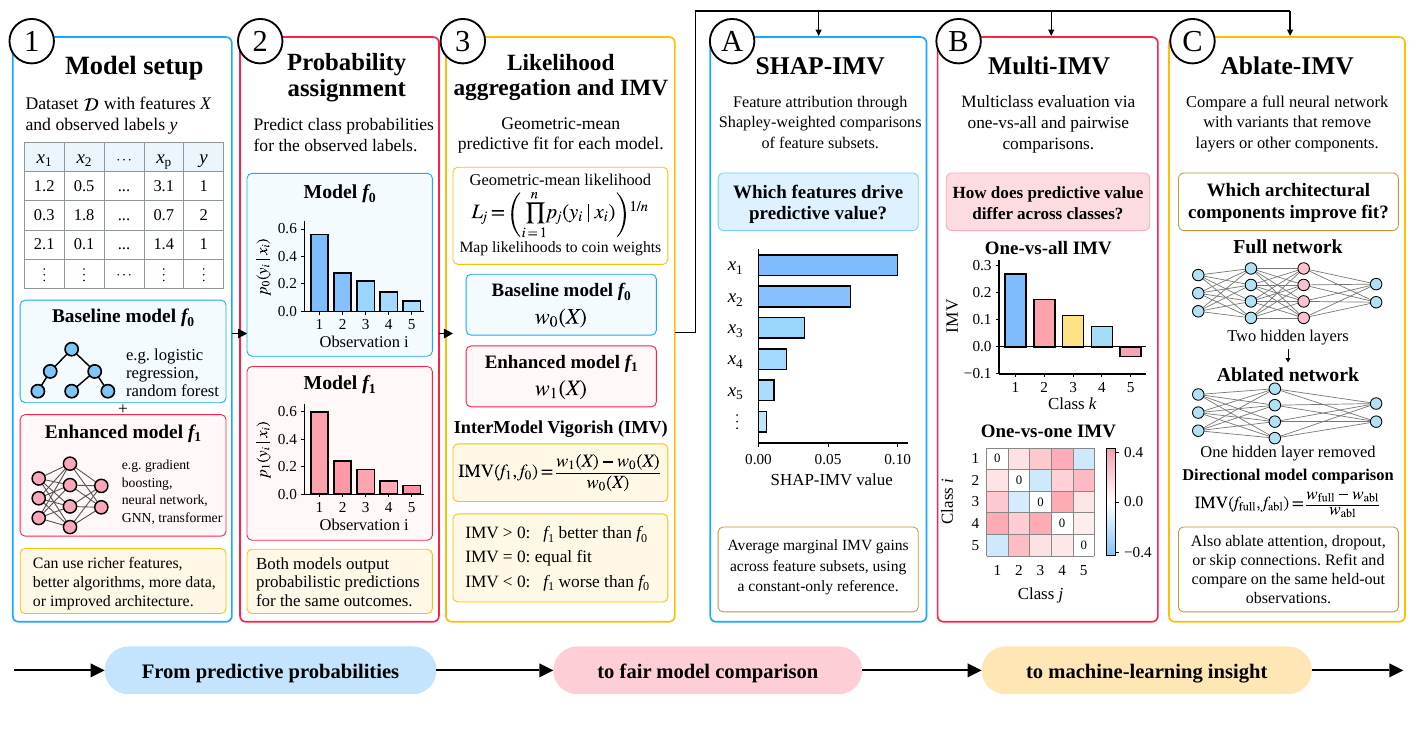}
    \caption{IMV calculation and applications. Steps \textbf{1--3} start with the probabilities that reference model $f_0$ and candidate model $f_1$ assign to the same observed outcomes. We aggregate these as geometric mean likelihoods. When both are at least $0.5$, Equation~\eqref{eq:coin-weight} gives equivalent coin weights $w_0$ and $w_1$. The directional score $\mathrm{IMV}(f_1,f_0)=(w_1-w_0)/w_0$ measures the relative change in coin weight; positive values indicate higher mean log-likelihood for $f_1$. (\textbf{A}) SHAP-IMV attributes gains in fit to features through Shapley-weighted marginal contributions from refitted subset models, using a constant-only reference. (\textbf{B}) Multi-IMV reuses multiclass predictions for one-vs-all and pairwise comparisons, renormalizing probabilities within each class pair. (\textbf{C}) Ablate-IMV compares full and refitted ablated models on the same held-out observations to assess architectural components or feature groups. Scores depend on the reference and evaluation data. All data, probabilities, scores, and networks shown are schematic.}
    \label{fig:Overview of IMV}
\end{figure*}

\subsection{The InterModel Vigorish (IMV)}
IMV interprets a difference in probabilistic fit through an equivalent betting problem~\cite{domingue2021intermodel,zhangintermodel,domingue2024intermodel}. For $L(p;D)\geq0.5$, the equivalent coin weight $w\in[0.5,1]$ matches the coin's expected log-likelihood to the model's mean log-likelihood:
\begin{equation}
 \log L(p; D) = w \log w + (1 - w) \log (1 - w).
 \label{eq:coin-weight}
\end{equation}
Here $\log L=N^{-1}\sum_i \ell_i$. We extend $x\log x$ continuously to $x=0$ by defining $0\log0:=0$. The right-hand side ranges from $-\log 2$ to zero; consequently, no exact coin weight exists when $L<0.5$.

We use the default settings of \texttt{imvpy} 1.2.0~\cite{imvpy2026}. Probabilities are clipped to $[10^{-9},1-10^{-9}]$, and below-chance likelihoods receive the boundary weight $w=0.5$ if $-\log(2L)\leq0.5$ nats. Larger deficits produce a warning and an undefined score. The boundary rule is an implementation convention rather than an exact solution of Equation~\eqref{eq:coin-weight}, so distinct poor predictions can receive the same weight. 

Consider models $f_0$ and $f_1$ with coin weights $w_0$ and $w_1$. Bettor A stakes one dollar on heads and B stakes $1/O$ dollars on tails. Fair odds for the baseline coin require
\begin{equation}
 w_0 \times \frac{1}{O} = (1 - w_0) \times 1.
\end{equation}
Thus $1/O=(1-w_0)/w_0$. If the coin is replaced by one with weight $w_1$ and the odds are held fixed, A's expected profit becomes
\begin{equation}
 w_1 \times \frac{1 - w_0}{w_0} - (1 - w_1) \times 1 = \frac{w_1 - w_0}{w_0}.
 \label{IMV}
\end{equation}
This betting edge, or `vigorish', defines IMV~\cite{domingue2021intermodel} (Figure~\ref{fig:Overview of IMV}). Positive values favor $f_1$ and negative values favor baseline $f_0$. The comparison requires predictions for the same outcomes, in the same order. With the baseline and evaluation set held fixed, IMV orders predictions identically to mean log-likelihood, and inversely to log loss, within the exact coin domain. Its upper bound is $1/w_0-1$, so the scale of possible improvement depends on the baseline and outcome prevalence.

\subsection{Calculating IMV in Python}
Box~\ref{cb:vanilla-imv} illustrates an IMV calculation with our \href{https://pypi.org/project/imvpy/1.2.0/}{\texttt{imvpy}} library. We simulate 1,000 binary outcomes from one predictor, fit logistic regression on 70\% of the observations, and evaluate its predictions on the remaining 30\%. For comparison, the baseline assigns the training-set prevalence to every test observation. We pass the baseline probabilities, enhanced probabilities, and observed outcomes to \texttt{vanilla\_imv}, in that order.

\begin{imvcodeblock}{Vanilla IMV}{cb:vanilla-imv}
\begin{lstlisting}[style=imvpython]
# Install once: python -m pip install imvpy
import numpy as np
from imvpy import vanilla_imv
from sklearn.linear_model import LogisticRegression
from sklearn.model_selection import train_test_split

# 1. Simulate a binary outcome from one predictor.
rng = np.random.default_rng(42)
x = rng.normal(size=(1000, 1))
p = 1 / (1 + np.exp(-(2 * x[:, 0] - 1)))
y = rng.binomial(1, p)

# 2. Fit on training data and predict test outcomes.
x_train, x_test, y_train, y_test = train_test_split(
    x, y, test_size=0.3, random_state=42)
model = LogisticRegression().fit(x_train, y_train)
p_enhanced = model.predict_proba(x_test)[:, 1]

# 3. Compare against the training prevalence.
p_baseline = y_train.mean()
imv = vanilla_imv(p_baseline, p_enhanced, y_test)
print(f'IMV: {imv:.3f}')
\end{lstlisting}
\end{imvcodeblock}

\noindent The resulting IMV is approximately 0.282. This places the improvement in probabilistic fit on an immediately interpretable scale: if the betting odds are set so that the prevalence baseline would break even, replacing the baseline with the fitted model yields an expected profit of approximately \$0.282 for each \$1 staked. Thus, rather than expressing the improvement only as an abstract difference in likelihood or log-likelihood, IMV translates it into the more intuitive quantity of an expected advantage. This provides a simple way to communicate the magnitude of the improvement and to compare models evaluated against the same reference model and evaluation data.

\section{Methodology}
\subsection{Overview}\label{sec:simulated-imv}
We use SHAP-IMV to attribute predictive value to features through Shapley weighting~\cite{SHAP}. Multi-IMV compares classes one-vs-all and one-vs-one~\cite{galar2011overview}, while Ablate-IMV compares architectural components and feature groups. Each comparison measures improvement relative to an explicit reference model. The examples below use simulated data and the \texttt{imvpy} library to demonstrate these three applications.

\subsection{SHAP-IMV for Feature Attribution}

Let $\mathcal{F}=\{1,\ldots,d\}$ index the features, and let $f_S$ use subset $S\subseteq\mathcal{F}$ plus a constant predictor. The reference $f()$ is the same estimator fitted on the constant alone. We define $v(S)$ as the mean held-out $\mathrm{IMV}(f_S,f())$ over cross-validation folds shared by all subsets, refitting both models on each fold's training observations. The empty-subset and reference models coincide, so $v(\varnothing)=0$. The contribution of feature $j$ is its Shapley-weighted average marginal gain~\cite{shapley1953value,SHAP}:
\begin{equation}
 \phi_j=\sum_{S\subseteq\mathcal{F}\setminus\{j\}}
 \frac{|S|!(d-|S|-1)!}{d!}\bigl[v(S\cup\{j\})-v(S)\bigr].
\end{equation}
Including the empty subset ensures that contributions are averaged across all feature-entry orders. Before rounding, allocations satisfy Shapley efficiency, $\sum_j\phi_j=v(\mathcal{F})-v(\varnothing)$, as well as symmetry, dummy-feature, and additivity properties. These properties do not guarantee unbiased estimation, causal identification, or social fairness. The values depend on the estimator, representation, reference, and evaluation data. 
Evaluating every subset requires $2^d$ fits per fold, which motivates our
eight-predictor specifications. Model fitting can therefore dominate the cost
of IMV scoring. The library rounds contributions to three decimals, so
displayed sums agree with $v(\mathcal{F})$ only within rounding precision. Section~\ref{Computing the SHAP-IMV in Python: A Toy Example} gives a simulated example.

\subsubsection{Simulated SHAP-IMV example}\label{Computing the SHAP-IMV in Python: A Toy Example}

Box~\ref{cb:shap-imv} generates 1,000 binary outcomes from two predictors. It uses \texttt{BinaryIMV} to evaluate every feature subset with logistic regression over three stratified folds, comparing each subset with a constant-only model fitted on the training observations.

The global SHAP-IMV values attribute the predictive gain to the two features. This example returns 0.209 for \texttt{x1} and 0.056 for \texttt{x2}.

\subsection{Multi-IMV for Multiclass Evaluation}
\smallskip\subsubsection{One-vs-All Comparisons}

Multi-IMV evaluates multiclass predictions through binary contrasts~\cite{bishop2006pattern,galar2011overview}. Within each fold, we fit one enhanced multiclass estimator $f_1$ and one constant-only reference $f_0$. Let $p_{rik}$ denote the probability assigned by model $r\in\{0,1\}$ to class $k$ for held-out instance $i$, with probability columns aligned by class label. For each class $k\in\{1,\ldots,K\}$, define

\begingroup
\mdfsetup{nobreak=true}
\begin{imvcodeblock}{SHAP-IMV}{cb:shap-imv}
\begin{lstlisting}[style=imvpython]
import numpy as np
import pandas as pd
from imvpy import BinaryIMV
from sklearn.linear_model import LogisticRegression

rng = np.random.default_rng(42)
x = rng.normal(size=(1000, 2))
p = 1 / (1 + np.exp(-(2 * x[:, 0] + x[:, 1] - 1)))
data = pd.DataFrame(x, columns=['x1', 'x2'])
data['y'] = rng.binomial(1, p)
evaluator = BinaryIMV(data=data, outcome_variable='y',
    optional_explanatory_variables=['x1', 'x2'],
    model_creator=lambda: LogisticRegression(),
    split_method='stratified_kfold', n_splits=3,
    random_seed=42)
evaluator.run_evaluation()
for feature in ['x1', 'x2']:
    value = evaluator.calculate_imvshapley_value(feature)
    print(f'SHAP-IMV for {feature}: {value:.3f}')
\end{lstlisting}
\end{imvcodeblock}
\endgroup

\begin{equation}
Y_i^{(k)} =
\begin{cases}
1 & \text{if } Y_i = k, \\
0 & \text{otherwise}.
\end{cases}
\end{equation}
We compare $p_{0ik}$ and $p_{1ik}$ with these binary outcomes on all held-out instances. This uses the original probability columns, without renormalization or separate binary classifiers. If $w_{0k}$ and $w_{1k}$ are the corresponding coin weights, then
\begin{equation}
\label{multiIMVcal}
 \mathrm{IMV}_k(f_1,f_0)=\frac{w_{1k}-w_{0k}}{w_{0k}}.
\end{equation}
Each class has its own reference weight, reflecting its binary prevalence and the reference probabilities on the evaluation fold.

\smallskip\subsubsection{The One-vs-One Multi-IMV Confusion Matrix}
Pairwise evaluation focuses on discrimination between two classes. It supplements one-vs-all evaluation but does not eliminate class imbalance. For classes $k\ne l$, retain only instances with $Y_i\in\{k,l\}$ and define
\begin{equation}
Y_i^{(k,l)} =
\begin{cases}
1 & \text{if } Y_i = k, \\
0 & \text{if } Y_i = l.
\end{cases}
\end{equation}
For both models, renormalize the retained class probabilities as $q_{rik}^{(k,l)}=p_{rik}/(p_{rik}+p_{ril})$. We reuse the fitted multiclass estimators without training a model for each class pair. Let $w_{r,kl}$ be the coin weight obtained from these probabilities and outcomes, and define $\lambda_{kl}=(w_{1,kl}-w_{0,kl})/w_{0,kl}$. For $K=3$, the (symmetric) Multi-IMV matrix is
\begin{equation}
 \mathbf{\Lambda}=\begin{pmatrix}
 0 & \lambda_{12} & \lambda_{13} \\
 \lambda_{12} & 0 & \lambda_{23} \\
 \lambda_{13} & \lambda_{23} & 0
 \end{pmatrix}.
 \label{multiIMVconfusion}
\end{equation}
Swapping the two class labels complements both the binary outcomes and probabilities, leaving their likelihoods unchanged. Thus; $\lambda_{kl}=\lambda_{lk}$, and we set the diagonal to zero by convention. Despite its name, this matrix contains baseline-relative probabilistic scores rather than confusion counts. We average both one-vs-all and pairwise scores across folds within each seed, then across seeds, and recommend reporting both alongside conventional metrics. Section~\ref{Computing the Multi-IMV Confusion Matrix in Python: A Toy Example} gives a simulated example.

\smallskip\subsubsection{Reading One-vs-All and Pairwise Scores}
The two views answer different questions. $\mathrm{IMV}_k$ is a class-specific diagnostic: it measures how much the features improve the probability of membership in class $k$, with all other classes pooled. $\lambda_{kl}$ is a pairwise diagnostic: among instances from classes $k$ and $l$ only, it measures how much the features help separate the two. The views use different populations and reference weights, and neither is an average of the other.

Both scores are bounded by class balance. If the reference assigns the binary prevalence $\pi$ to every instance, then $\log L=\pi\log\pi+(1-\pi)\log(1-\pi)$, and Equation~\eqref{eq:coin-weight} gives $w_0=\max(\pi,1-\pi)$. The bound $1/w_0-1$ is then the minority-to-majority ratio of the contrast: $\pi_k/(1-\pi_k)$ for one-vs-all with $\pi_k<0.5$, and $\min(n_k,n_l)/\max(n_k,n_l)$ for a pair with class counts $n_k$ and $n_l$. A frequent class therefore has a high one-vs-all bound but low pairwise bounds against rarer classes, whereas a class of intermediate frequency can have high pairwise bounds throughout. 

We therefore compare the same class or pair across models directly, and compare different classes or pairs relative to their bounds. The ratio $(w_1-w_0)/(1-w_0)$ gives the share of the attainable improvement achieved. Values near one indicate near-complete separation, while pairs well below one show where classes remain confused.

\subsubsection{Simulated Multi-IMV example}\label{Computing the Multi-IMV Confusion Matrix in Python: A Toy Example}

Box~\ref{cb:multi-imv} generates 900 observations with three outcome classes. It uses \texttt{MulticlassIMV} to compare logistic regression predictions with a constant-only model over three stratified folds.

\noindent The first output is the symmetric pairwise Multi-IMV matrix, with classes indexing its rows and columns and zeros on the diagonal. The second compares each class with all remaining classes. Both report mean IMV over the held-out folds.

\begingroup
\mdfsetup{nobreak=true}
\begin{imvcodeblock}{Multi-IMV}{cb:multi-imv}
\begin{lstlisting}[style=imvpython]
import pandas as pd
from imvpy import MulticlassIMV
from sklearn.datasets import make_classification
from sklearn.linear_model import LogisticRegression

X, y = make_classification(
    n_samples=900, n_features=3, n_informative=3,
    n_redundant=0, n_classes=3, n_clusters_per_class=1,
    class_sep=0.5, flip_y=0, random_state=42)
features = ['x1', 'x2', 'x3']
data = pd.DataFrame(X, columns=features).assign(y=y)

multi = MulticlassIMV(data=data, outcome_variable='y',
    optional_explanatory_variables=features,
    model_creator=lambda: LogisticRegression(max_iter=2000),
    n_splits=3, random_state=42, stratified=True)
_, pairwise = multi.k_fold_imv_matrix()
_, one_vs_rest = multi.k_fold_one_vs_all()
print('Pairwise IMV:')
print(pairwise.round(3))
print('One-vs-rest IMV:')
print(pd.Series(one_vs_rest, index=pairwise.index).round(3))
\end{lstlisting}
\end{imvcodeblock}
\endgroup

\subsection{Ablate-IMV for Model Component Comparisons}
Ablation studies typically compare full models with variants using accuracy and other performance metrics~\cite{vaswani2017attention,he2019rethinking}. Ablate-IMV measures each variant's probabilistic fit against a reference and compares variants directly. Consider a full model $F_0$, ablations $F_1$ and $F_2$, and an uninformative reference $F()$ predicting 50\% for every binary outcome ($w()=0.5$). Equation~\eqref{IMV} gives
\begin{equation}
\text{IMV}(F_i, F()) = \frac{w_i - w()}{w()}, \quad \text{where } i \in \{0, 1, 2\}.
\end{equation}
\noindent These scores measure fit relative to the 0.5-probability reference, not the original model. Comparing two ablations directly gives
\begin{equation}
\text{IMV}(F_i, F_j) = \frac{w_i - w_j}{w_j}, \quad \text{where } i, j \in \{1, 2\}, \, i \neq j.
\end{equation}

\noindent We can also compare the original model with each ablation:
\begin{equation}
\begin{aligned}
\text{IMV}(F_0, F_i)
= \frac{w_0 - w_i}{w_i},
\quad \text{where } i \in \{1, 2\}.
\end{aligned}
\end{equation}
In the directional $\mathbf{M}$, rows represent enhanced models and columns represent basic models:
\begin{equation}
 \mathbf{M}=\begin{pmatrix}
0.0 & \text{IMV}(F_0, F_1) & \text{IMV}(F_0, F_2) \\
\text{IMV}(F_1, F_0) & 0.0 & \text{IMV}(F_1, F_2) \\
\text{IMV}(F_2, F_0) & \text{IMV}(F_2, F_1) & 0.0
\end{pmatrix}.
\end{equation}
Each column's reference weight supplies the denominator, so $\mathbf{M}$ is generally asymmetric. Positive entries favor the row model, and reversing a comparison need not negate its IMV. Reporting both $\text{IMV}(F_i,F())$ and $\mathbf{M}$ separates fit against an uninformative null from directional comparisons among models.
\subsubsection{Simulated Ablate-IMV example}\label{sec:simulated-ablate-imv}

Box~\ref{cb:ablate-imv} simulates a binary outcome and fits two neural networks: one with two hidden layers and one with the second hidden layer removed. Both are fitted from scratch on the same training observations. We then use \texttt{AblationIMV} to compare their probabilities on the same test observations.

\begingroup
\mdfsetup{nobreak=true}
\begin{imvcodeblock}{Ablate-IMV}{cb:ablate-imv}
\begin{lstlisting}[style=imvpython]
import numpy as np
import pandas as pd
from imvpy import AblationIMV
from sklearn.model_selection import train_test_split
from sklearn.neural_network import MLPClassifier

rng = np.random.default_rng(42)
X = rng.normal(size=(2000, 3))
p = 1 / (1 + np.exp(-(2 * X[:, 0] + X[:, 1])))
y = rng.binomial(1, p)
X_train, X_test, y_train, y_test = train_test_split(
    X, y, test_size=0.3, stratify=y, random_state=42)

predictions = {}
variants = {'Full': (8, 8), 'Ablated': (8,)}
for name, layers in variants.items():
    model = MLPClassifier(hidden_layer_sizes=layers,
        solver='lbfgs', alpha=1, max_iter=2000,
        random_state=42).fit(X_train, y_train)
    predictions[name] = pd.DataFrame({'True Label': y_test,
        'Positive Probability': model.predict_proba(
            X_test)[:, 1]})

matrix = AblationIMV.calculate_imv_matrix(predictions)
print(matrix.round(3))
\end{lstlisting}
\end{imvcodeblock}
\endgroup

\noindent Each matrix entry compares the row model with the column model. The negative \texttt{Full}--\texttt{Ablated} entry in this example indicates that the additional layer does not improve held-out fit.

\section{Experiments}
\subsection{Design and Implementation}
\smallskip\subsubsection{Datasets}We examine SHAP-IMV on OpenML Titanic~\cite{titanic_openml_40945}, Wisconsin Diagnostic Breast Cancer~\cite{breast_cancer_wisconsin_diagnostic_17}, and Adult Income~\cite{adult_2}, restricting each specification to eight predictors so that exact enumeration remains tractable. The multiclass experiments use Nursery~\cite{nursery_76}, Car Evaluation~\cite{car_evaluation_19}, and Dry Bean~\cite{dry_bean_602} from the UCI Machine Learning Repository~\cite{asuncion2007uci}. For ablation, the tasks are MNIST odd/even digit classification~\cite{mnist_openml_554}, UCI HAR upstairs/downstairs activity recognition~\cite{har_smartphones_240}, and MAGIC Gamma gamma/hadron classification~\cite{magic_gamma_159}. The GNN case study uses PROTEINS, NCI1, NCI109, Mutagenicity, AIDS, and DD~\cite{morris2020tudataset}. Sections~\ref{datasets} and~\ref{datasets:gnn} describe the samples and preprocessing.

\smallskip\subsubsection{Comparison Methods}We compare SHAP-IMV rankings with SHAP~\cite{SHAP}, LIME~\cite{LIME}, and Anchors~\cite{Anchors}, and report Multi-IMV alongside per-class Brier loss~\cite{brier1950verification}, accuracy, precision, and recall~\cite{powers2020evaluation}. Section~\ref{BaselineAlgorithms} defines the methods and metrics.

\smallskip\subsubsection{Models and Training}For SHAP-IMV, logistic regression allows up to 5{,}000 iterations~\cite{pedregosa2011scikit}; XGBoost~\cite{chen2016xgboost} and LightGBM~\cite{ke2017lightgbm} each use 60 boosting iterations, maximum tree depth three, and learning rate 0.2. Exact attribution uses five-fold stratified cross-validation for each of ten seeds (42--51); we report the mean and sample standard deviation of $\phi_j$ across seeds, with seed-level contributions rounded to three decimals before averaging. We then evaluate the selected top-five subsets using ten-fold stratified cross-validation with the same ten seeds. Standardization uses the full dataset, and feature selection precedes the refit evaluation folds. Multi-IMV evaluates logistic regression, XGBoost, and LightGBM with five-fold stratified cross-validation across ten seeds (42--51).

For Ablate-IMV, the MNIST model is a two-convolution CNN trained for one epoch with Adadelta (learning rate 1.0, batch size 256). UCI HAR uses a two-layer bidirectional GRU with attention, trained for 15 epochs with Adam (learning rate $10^{-3}$, batch size 128); MAGIC Gamma uses an MLP with hidden widths 64 and 32, trained for 12 epochs with AdamW (learning rate $10^{-3}$, batch size 256). Each full and ablated model is evaluated for ten seeds (42--51). MNIST and HAR retain their official partitions, while MAGIC Gamma uses seed-specific stratified 75/25 splits. 
The GNN case study uses a three-layer GCN with hidden width 64, dropout 0.5, learning rate 0.001, and batch size 512. Training lasts 200 epochs for each of ten seeds, with seed-specific stratified 70/15/15 training/validation/test splits. We report held-out metrics at every epoch, including the untrained epoch 0.

\smallskip\subsubsection{Software and Hardware} The experiments use \texttt{imvpy} 1.2.0, available on PyPI~\cite{imvpy2026}. The saved GNN manifests record CPU execution with PyTorch 2.9.0~\cite{paszke2019pytorch}, PyTorch Geometric 2.7.0~\cite{Fey/Lenssen/2019}, NumPy 1.26.4, and Scikit-learn 1.7.2~\cite{pedregosa2011scikit}. The runtime benchmark used Python 3.12.7 and an Intel Core i9-14900KF CPU. The notebooks record other dependencies and available details of the runs, as described in our {}\texttt{Code and data availability} statement.

\subsection{Datasets and Preprocessing}\label{datasets}

\subsubsection{Datasets for SHAP-IMV}
Each binary classification task uses eight specified predictors: passenger characteristics for Titanic, cellular measurements for Breast Cancer, and socioeconomic characteristics for Adult Income.

\paragraph{Titanic}
The OpenML Titanic dataset (ID 40945, version 1) contains 1{,}309 passengers, including 500 survivors~\cite{titanic_openml_40945}. We code survival as the positive target outcome and use sex, title extracted from the passenger's name, ticket class, age multiplied by ticket class, fare, embarkation port, an indicator for travelling without siblings, spouses, parents, or children aboard, and age itself. Titles are grouped into five categories, with a separate code for unmatched names. We impute the 263 missing ages and one missing fare with their medians, assign the two missing embarkation ports to Southampton, and retain all passengers.

\paragraph{Breast Cancer Wisconsin (Diagnostic)}
The diagnostic dataset contains 569 breast-mass target samples: 212 malignant and 357 benign~\cite{breast_cancer_wisconsin_diagnostic_17}. Its 30 continuous measurements describe cell nuclei in digitized images of fine-needle aspirates. We select mean radius, texture, smoothness, compactness, symmetry, fractal dimension, concavity, and concave points, and code malignancy as one. All eight inputs are complete, so no observations are removed or imputed.

\paragraph{Adult Income}
This dataset contains 48{,}842 records extracted from the 1994 Census database, of which 11{,}687 have annual income above \$50{,}000~\cite{adult_2}. The selected inputs are age, educational attainment (\texttt{education-num}), weekly working hours, capital gains, capital losses, and indicators for female sex, married status, and white race. The positive outcome is income above the threshold. We harmonize labels and pool the original training and test files for cross-validation rather than using the official test partition. Missing values occur in other source fields, but the eight selected inputs are complete, so all records are retained.

\subsubsection{Datasets for Multi-IMV}
\label{Datasets for Multi-IMV}
The multiclass benchmarks differ in their class frequencies. We retain these frequencies within the selected classes for both one-versus-rest and pairwise comparisons.

\paragraph{Nursery}
This dataset comprises 12{,}960 cases generated from a hierarchical decision model for ranking nursery-school applications~\cite{nursery_76} as the target feature. The eight categorical predictors describe parental circumstances, childcare arrangements, family structure, housing, finances, and social and health conditions. We exclude the \texttt{recommend} and \texttt{very\_recom} classes, which contain two and 328 cases, respectively. The retained sample has 12{,}630 observations: 4{,}320 not recommended, 4{,}266 priority, and 4{,}044 special priority.

\paragraph{Car Evaluation}
This dataset contains 1{,}728 configurations generated by a hierarchical acceptability model~\cite{car_evaluation_19}. Its six categorical predictors describe purchase and maintenance costs, doors, passenger capacity, luggage space, and safety. We retain all four outcome classes: unacceptable (1{,}210), acceptable (384), good (69), and very good (65).

\paragraph{Dry Bean}
This dataset contains 13{,}611 photographed grains represented by 16 size and shape measurements~\cite{dry_bean_602}. We retain all seven varieties as classes: Dermason (3{,}546), Sira (2{,}636), Seker (2{,}027), Horoz (1{,}928), Cali (1{,}630), Barbunya (1{,}322), and Bombay (522).\\

None of the three multiclass datasets requires imputation. Nursery and Car inputs are coded as integer categories, and Dry Bean measurements are standardized. Comparisons between estimators depend on these choices, including the use of ordinal codes for nominal predictors.

The SHAP-IMV and Multi-IMV analyses use five stratified folds across ten seeds. Titanic imputation and SHAP-IMV and Dry Bean standardization are performed before constructing these folds.

\subsubsection{Datasets for Ablate-IMV}
We ablate models for images, sensor sequences, and tabular data, evaluating every variant on the same test observations within each dataset and seed.

\paragraph{MNIST}
This canonical dataset contains 70{,}000 grayscale images of handwritten digits, each represented by $28\times28$ pixels~\cite{mnist_openml_554}. We use OpenML \texttt{mnist\_784}, version 1, preserving the canonical 60{,}000-image training and 10{,}000-image test partitions. Odd digits are coded as positive classes, and even ones as negative. Pixels are divided by 255 and normalized using the fixed mean 0.1307 and standard deviation 0.3081. We use this parity task to compare a convolutional network with its architectural ablations.

\paragraph{Human Activity Recognition Using Smartphones}
UCI HAR records six activities performed by 30 participants carrying waist-mounted smartphones~\cite{har_smartphones_240}. We retain walking upstairs (one) and downstairs (zero). The supplied participant-disjoint partition yields 2{,}059 training windows from 21 participants and 891 test windows from nine participants. Each window spans 2.56 seconds and contains 128 readings from nine channels: three axes each of body acceleration, total acceleration, and angular velocity. Channel normalization uses training windows alone. These sequences are used for recurrent-network ablations.

\paragraph{MAGIC Gamma Telescope}
This dataset contains 19{,}020 simulated telescope events with ten continuous image measurements and no missing predictors~\cite{magic_gamma_159}. Gamma events are coded one (12{,}332 cases) and hadron backgrounds zero (6{,}688). Stratified 75/25 splits provide 14{,}265 training and 4{,}755 test observations per seed, with standardization fitted on the training observations. We hold the multilayer perceptron's architecture fixed and remove groups of measurements describing geometry, size and concentration, moments, or orientation and distance. The simulated class proportions do not represent gamma-event prevalence in observational telescope data.

\subsection{Comparison Methods and Metrics}\label{BaselineAlgorithms}
\subsubsection{Feature Explanation Methods}
For each local explanation method, the estimator is fitted to all eight standardized predictors on the full analysis sample. We sample rows to explain without replacement for each seed, using all rows when the sample is smaller than the explanation budget. Importance scores are averaged over seeds to select five predictors. These methods explain a fixed fitted model, whereas SHAP-IMV refits subsets within cross-validation; the explanation scales and computational budgets also differ.

\paragraph{SHAP}
SHAP allocates a prediction among features relative to a background distribution~\cite{SHAP,strumbelj2010efficient}. We average absolute contributions over at most 2{,}000 rows per seed. For logistic regression, \texttt{LinearExplainer} uses interventional feature masking: we supply 200 background rows, which SHAP's default independent masker subsamples to 100. XGBoost uses its native exact TreeSHAP \texttt{pred\_contribs} interface, excluding the bias column. LightGBM uses \texttt{TreeExplainer} without a supplied background, giving tree-path-dependent perturbation. The resulting contributions are in the models' raw-output units rather than IMV units. Their interpretation depends on assumptions about correlated predictors and does not establish causal effects.

\paragraph{LIME}
LIME fits a distance-weighted local surrogate to model predictions on perturbed inputs~\cite{LIME,garreau2020explaining}. We explain at most 200 rows per seed, using 2{,}000 perturbations per row and requesting all eight features and the probability of class one. Continuous predictors are discretized into quartiles; binary predictors are declared categorical. We rank features by their mean absolute surrogate coefficient across explained rows. This score measures coefficient magnitude rather than a signed contribution to predictive fit.

\paragraph{Anchors}
Anchors searches for feature rules that preserve a predicted class under its perturbation distribution~\cite{Anchors}. We use Alibi's \texttt{AnchorTabular} on at most 100 rows per seed, with a target precision of 0.95 and quartile cut points at 25, 50, and 75 percent for all inputs. We retain returned rules even when the search misses its precision target. Failed searches contribute zero counts but remain in the denominator. Because achieved precision and failure counts were not saved, we cannot verify that the returned rules met the target. For feature $j$, let $c_j$ count its rule inclusions and let $r_j$ sum its reciprocal positions within returned rules (first position one). With $n$ attempted explanations and $c_{\max}=\max_j c_j$, the seed-level ranking score is
\[
 s_j=\frac{c_j}{n}+\frac{r_j}{2n\max(c_{\max},1)}.
\]
The second term breaks count ties within a seed. Averaging over seeds therefore gives a frequency score adjusted for rule position, rather than a ranking based on frequency alone. The comparisons are descriptive and use unequal explanation budgets, so they cannot establish that one method is uniformly superior. Assessing how well the selection procedures generalize would require nested feature selection and preprocessing fitted within each fold.

\subsubsection{Multiclass Evaluation Metrics}

\paragraph{Brier Score Loss}
Brier loss measures squared probability error~\cite{brier1950verification}. For binary outcomes,

\[
\text{Brier Score} = \frac{1}{N} \sum_{i=1}^{N} (y_i - \hat{p}_i)^2,
\]

\noindent where $N$ counts instances, $y_i\in\{0,1\}$ is the observed label, and $\hat{p}_i$ is its predicted positive-class probability.

\paragraph{Accuracy, Precision, and Recall}
Accuracy, precision, and recall summarize classification decisions~\cite{Accuracy,murphy2012machine}. Let TP and TN count correct positive and negative predictions, and FP and FN count false positives and negatives. Accuracy is the fraction correctly classified:
\[
\text{Accuracy} = \frac{TP + TN}{TP + TN + FP + FN}.
\]
Precision is the fraction of positive predictions that are correct:
\[
\text{Precision} = \frac{TP}{TP + FP}.
\]
Recall is the fraction of actual positives recovered:
\[
\text{Recall} = \frac{TP}{TP + FN}.
\]

\subsection{SHAP-IMV Results}

\begin{figure*}[h]
    \centering
    \includegraphics[width=1\linewidth]{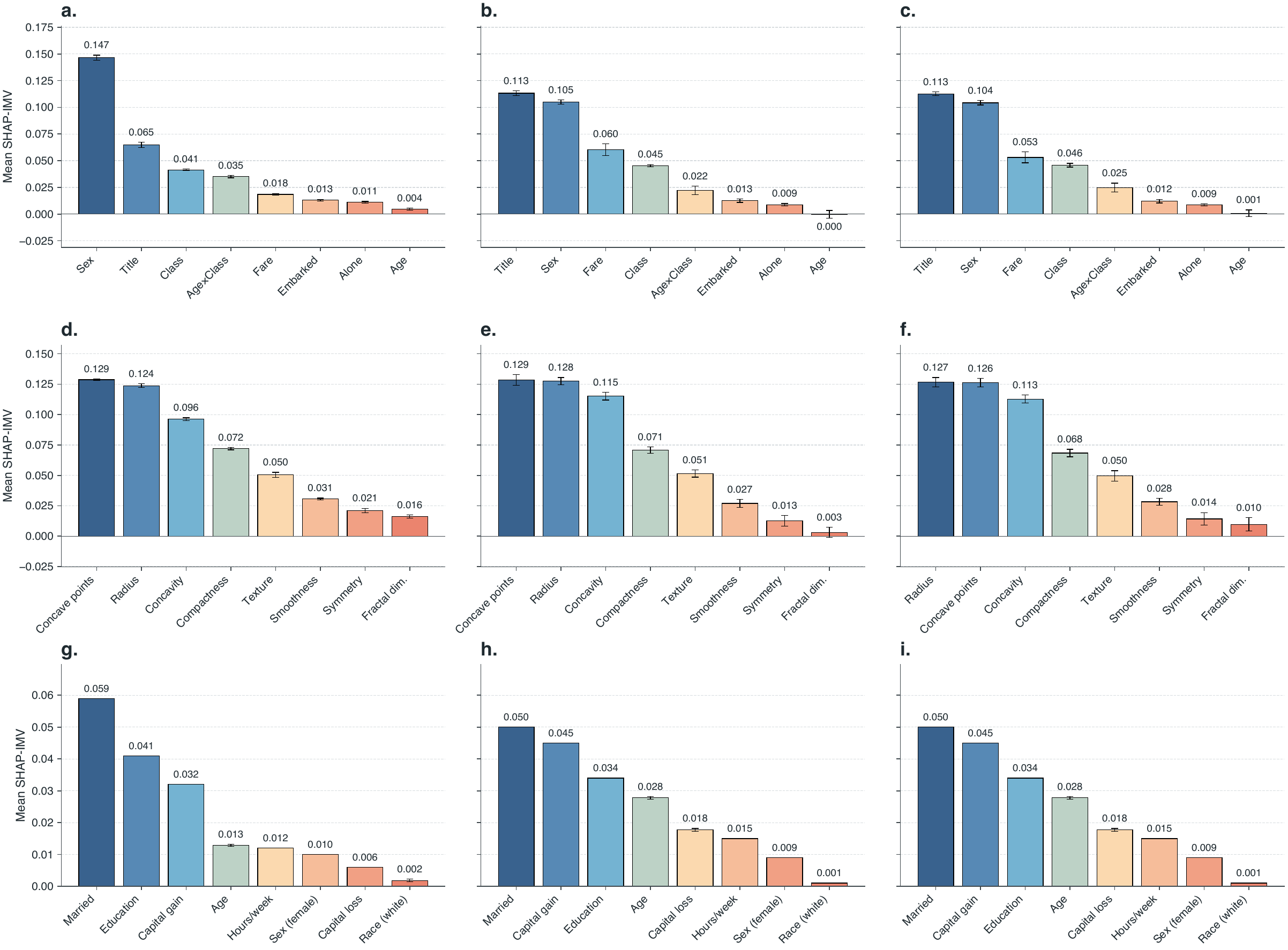}
    \caption{SHAP-IMV feature contributions for Titanic (\textbf{a.}--\textbf{c.}), Wisconsin Diagnostic Breast Cancer (\textbf{d.}--\textbf{f.}), and Adult Income (\textbf{g.}--\textbf{i.}). Columns show logistic regression, XGBoost, and LightGBM; each panel ranks eight predictors by mean signed SHAP-IMV. All 256 subsets are evaluated by five-fold stratified cross-validation against the same estimator fitted to a constant input. Shapley contributions use the mean IMV across folds. Bars and labels give means across ten seeds (42--51), with whiskers spanning one sample standard deviation rather than confidence intervals. Standardization uses the full dataset before cross-validation. Y-scales are shared within rows, and the contributions describe predictive fit rather than causal effects.}
    \label{fig:SHAPIMV_Titanic}
\end{figure*}

\begin{table*}[tp]
\centering
\footnotesize
\tabcolsep=3pt
    \begin{tabular}{cccccccccc}
        \toprule
         \multicolumn{2}{c}{\multirow{2}{*}{Methods}} & \multicolumn{2}{c}{Titanic} & \multicolumn{2}{c}{Breast Cancer} & \multicolumn{2}{c}{Adult Income} & \multicolumn{2}{c}{Average}\\
         &  & Accuracy & Precision & Accuracy & Precision & Accuracy & Precision & Accuracy & Precision \\
        \midrule
        \multirow{4}{*}{LR}
        & SHAP-IMV & \underline{79.50} & \textbf{74.36} & \underline{93.92} & \underline{94.27} & \underline{83.78} & \underline{70.94} & 85.73 & \underline{79.85}\\
        & \cellcolor{gray!15}SHAP & \cellcolor{gray!15}\textbf{79.74} & \cellcolor{gray!15}\underline{73.95} & \cellcolor{gray!15}\textbf{94.10} & \cellcolor{gray!15}\textbf{94.46} & \cellcolor{gray!15}\underline{83.78} & \cellcolor{gray!15}\underline{70.94} & \cellcolor{gray!15}\underline{85.87} & \cellcolor{gray!15}79.78\\
        & LIME & \textbf{79.74} & \underline{73.95} & \textbf{94.10} & \textbf{94.46} & \textbf{84.16} & \textbf{71.35} & \textbf{86.00} & \textbf{79.92}\\
        & \cellcolor{gray!15}Anchors & \cellcolor{gray!15}\underline{79.50} & \cellcolor{gray!15}\textbf{74.36} & \cellcolor{gray!15}\underline{93.92} & \cellcolor{gray!15}\underline{94.27} & \cellcolor{gray!15}\underline{83.78} & \cellcolor{gray!15}\underline{70.94} & \cellcolor{gray!15}85.73 & \cellcolor{gray!15}\underline{79.85}\\
        \midrule
        \multirow{4}{*}{XGB}
        & SHAP-IMV & \textbf{81.16} & \underline{79.33} & \underline{93.76} & \underline{93.11} & \underline{85.97} & \textbf{79.42} & \textbf{86.96} & \underline{83.95}\\
        & \cellcolor{gray!15}SHAP & \cellcolor{gray!15}\textbf{81.16} & \cellcolor{gray!15}\underline{79.33} & \cellcolor{gray!15}\textbf{94.22} & \cellcolor{gray!15}\textbf{93.34} & \cellcolor{gray!15}85.29 & \cellcolor{gray!15}75.77 & \cellcolor{gray!15}\underline{86.89} & \cellcolor{gray!15}82.81\\
        & LIME & \underline{80.44} & \textbf{80.47} & \textbf{94.22} & \textbf{93.34} & \textbf{85.98} & \underline{78.96} & 86.88 & \textbf{84.25}\\
        & \cellcolor{gray!15}Anchors & \cellcolor{gray!15}\textbf{81.16} & \cellcolor{gray!15}\underline{79.33} & \cellcolor{gray!15}\textbf{94.22} & \cellcolor{gray!15}\textbf{93.34} & \cellcolor{gray!15}85.29 & \cellcolor{gray!15}75.77 & \cellcolor{gray!15}\underline{86.89} & \cellcolor{gray!15}82.81\\
        \midrule
        \multirow{4}{*}{LGBM}
        & SHAP-IMV & \textbf{81.00} & \textbf{79.48} & \underline{93.97} & \underline{93.41} & \textbf{86.00} & \textbf{79.45} & \textbf{86.99} & \textbf{84.11}\\
        & \cellcolor{gray!15}SHAP & \cellcolor{gray!15}\textbf{81.00} & \cellcolor{gray!15}\textbf{79.48} & \cellcolor{gray!15}\textbf{94.39} & \cellcolor{gray!15}\textbf{93.87} & \cellcolor{gray!15}\underline{85.30} & \cellcolor{gray!15}75.91 & \cellcolor{gray!15}\underline{86.90} & \cellcolor{gray!15}83.09\\
        & LIME & \underline{80.02} & \underline{79.34} & \textbf{94.39} & \textbf{93.87} & \textbf{86.00} & \underline{79.03} & 86.80 & \underline{84.08}\\
        & \cellcolor{gray!15}Anchors & \cellcolor{gray!15}\textbf{81.00} & \cellcolor{gray!15}\textbf{79.48} & \cellcolor{gray!15}\textbf{94.39} & \cellcolor{gray!15}\textbf{93.87} & \cellcolor{gray!15}\underline{85.30} & \cellcolor{gray!15}75.91 & \cellcolor{gray!15}\underline{86.90} & \cellcolor{gray!15}83.09\\
        \bottomrule
    \end{tabular}
\caption{Predictive performance after top-5 feature selection by SHAP-IMV, SHAP, LIME, and Anchors. LR, XGB, and LGBM denote logistic regression, XGBoost, and LightGBM. Each dataset uses a fixed set of 8 variables drawn from the source data; Breast Cancer refers to the Wisconsin Diagnostic dataset. Rankings are averaged over seeds to select a fixed subset for refitting the same estimator. Accuracy and positive-class precision are percentages averaged over 10 stratified folds for each of 10 seeds; Average gives equal weight to the 3 datasets. Preprocessing and feature selection precede refit cross-validation, so the results describe these fixed subsets rather than a nested evaluation of the selection procedures. Within each estimator and column, bold and underlining mark the best and second-best distinct values at the displayed precision; ties share formatting, and columns in which all methods agree are unmarked.}
\label{SHAPIMVTabulartable}
\end{table*}

We compare SHAP-IMV with rankings obtained from mean absolute SHAP values~\cite{SHAP}, mean absolute LIME weights~\cite{LIME}, and anchor inclusion frequencies adjusted for rule position~\cite{Anchors}. After averaging importance scores over ten seeds, each method selects five of the eight predictors. We refit the same estimator on each fixed subset using ten-fold stratified cross-validation per seed and report mean accuracy and positive-class precision in Table~\ref{SHAPIMVTabulartable}. Since preprocessing and selection use the full dataset before refit evaluation, this comparison describes the selected subsets; it does not provide a nested estimate of how the selection procedures generalize. 

The four methods largely agree. SHAP-IMV selects the same five predictors as
at least one alternative in five of the nine dataset+estimator settings, and
shares at least four of five predictors with every alternative in all but one. Its accuracy is never more than 0.5 percentage points below the best alternative. For the boosting models on Adult Income, SHAP-IMV's subsets give the highest precision. With equal weight given to each dataset, SHAP-IMV selects the subsets with the highest mean accuracy for XGBoost (86.96\%) and LightGBM (86.99\%), while LIME leads for logistic regression (86.00\%). On Breast Cancer, however, SHAP-IMV subsets are slightly less accurate than SHAP or LIME subsets for every estimator. Methods that select identical sets share the same evaluation results. 


\begin{table*}[tp]
\centering
\footnotesize
\tabcolsep=3pt
    \begin{tabular}{cccccccccc}
        \toprule
         \multirow{2}{*}{Dataset} & \multirow{2}{*}{Model} & \multicolumn{8}{c}{Rank by mean SHAP-IMV}\\
         &  & 1 & 2 & 3 & 4 & 5 & 6 & 7 & 8 \\
        \midrule
        \multirow[t]{3}{*}{\shortstack{Titanic}}
        & LR & \shortstack{Sex\\(0.147)} & \shortstack{Title\\(0.065)} & \shortstack{Class\\(0.041)} & \shortstack{Age\,$\times$\,Class\\(0.035)} & \shortstack{Fare\\(0.018)} & \shortstack{Embarked\\(0.013)} & \shortstack{Alone\\(0.011)} & \shortstack{Age\\(0.004)}\\
        \addlinespace[2pt]
        & \cellcolor{gray!15}XGB & \cellcolor{gray!15}\shortstack{Title\\(0.113)} & \cellcolor{gray!15}\shortstack{Sex\\(0.105)} & \cellcolor{gray!15}\shortstack{Fare\\(0.060)} & \cellcolor{gray!15}\shortstack{Class\\(0.045)} & \cellcolor{gray!15}\shortstack{Age\,$\times$\,Class\\(0.022)} & \cellcolor{gray!15}\shortstack{Embarked\\(0.013)} & \cellcolor{gray!15}\shortstack{Alone\\(0.009)} & \cellcolor{gray!15}\shortstack{Age\\(-0.000)}\\
        \addlinespace[2pt]
        & LGBM & \shortstack{Title\\(0.113)} & \shortstack{Sex\\(0.104)} & \shortstack{Fare\\(0.053)} & \shortstack{Class\\(0.046)} & \shortstack{Age\,$\times$\,Class\\(0.025)} & \shortstack{Embarked\\(0.012)} & \shortstack{Alone\\(0.009)} & \shortstack{Age\\(0.001)}\\
        \addlinespace[2pt]
        \midrule
        \multirow[t]{3}{*}{\shortstack{Breast\\Cancer}}
        & LR & \shortstack{Concave points\\(0.129)} & \shortstack{Radius\\(0.124)} & \shortstack{Concavity\\(0.096)} & \shortstack{Compactness\\(0.072)} & \shortstack{Texture\\(0.050)} & \shortstack{Smoothness\\(0.031)} & \shortstack{Symmetry\\(0.021)} & \shortstack{Fractal dim.\\(0.016)}\\
        \addlinespace[2pt]
        & \cellcolor{gray!15}XGB & \cellcolor{gray!15}\shortstack{Concave points\\(0.129)} & \cellcolor{gray!15}\shortstack{Radius\\(0.128)} & \cellcolor{gray!15}\shortstack{Concavity\\(0.115)} & \cellcolor{gray!15}\shortstack{Compactness\\(0.071)} & \cellcolor{gray!15}\shortstack{Texture\\(0.051)} & \cellcolor{gray!15}\shortstack{Smoothness\\(0.027)} & \cellcolor{gray!15}\shortstack{Symmetry\\(0.013)} & \cellcolor{gray!15}\shortstack{Fractal dim.\\(0.003)}\\
        \addlinespace[2pt]
        & LGBM & \shortstack{Radius\\(0.127)} & \shortstack{Concave points\\(0.126)} & \shortstack{Concavity\\(0.113)} & \shortstack{Compactness\\(0.068)} & \shortstack{Texture\\(0.050)} & \shortstack{Smoothness\\(0.028)} & \shortstack{Symmetry\\(0.014)} & \shortstack{Fractal dim.\\(0.010)}\\
        \addlinespace[2pt]
        \midrule
        \multirow[t]{3}{*}{\shortstack{Adult\\Income}}
        & LR & \shortstack{Married\\(0.059)} & \shortstack{Education\\(0.041)} & \shortstack{Capital gain\\(0.032)} & \shortstack{Age\\(0.013)} & \shortstack{Hours/week\\(0.012)} & \shortstack{Sex (female)\\(0.010)} & \shortstack{Capital loss\\(0.006)} & \shortstack{Race (white)\\(0.002)}\\
        \addlinespace[2pt]
        & \cellcolor{gray!15}XGB & \cellcolor{gray!15}\shortstack{Married\\(0.050)} & \cellcolor{gray!15}\shortstack{Capital gain\\(0.045)} & \cellcolor{gray!15}\shortstack{Education\\(0.034)} & \cellcolor{gray!15}\shortstack{Age\\(0.028)} & \cellcolor{gray!15}\shortstack{Capital loss\\(0.018)} & \cellcolor{gray!15}\shortstack{Hours/week\\(0.015)} & \cellcolor{gray!15}\shortstack{Sex (female)\\(0.009)} & \cellcolor{gray!15}\shortstack{Race (white)\\(0.001)}\\
        \addlinespace[2pt]
        & LGBM & \shortstack{Married\\(0.050)} & \shortstack{Capital gain\\(0.045)} & \shortstack{Education\\(0.034)} & \shortstack{Age\\(0.028)} & \shortstack{Capital loss\\(0.018)} & \shortstack{Hours/week\\(0.015)} & \shortstack{Sex (female)\\(0.009)} & \shortstack{Race (white)\\(0.001)}\\
        \addlinespace[2pt]
        \bottomrule
    \end{tabular}
\caption{Complete SHAP-IMV feature rankings for the three binary classification examples. LR, XGB, and LGBM denote logistic regression, XGBoost, and LightGBM. Each dataset uses a fixed set of 8 variables drawn from the source data; Breast Cancer refers to the Wisconsin Diagnostic dataset. Predictors are ordered by decreasing mean exact Shapley contribution to model IMV; parentheses give means to three decimals. Attribution uses 5-fold stratified cross-validation for each of 10 seeds. The 10-fold refit evaluation in Table~\ref{SHAPIMVTabulartable} is separate and evaluates the fixed subset in ranks 1--5 for its SHAP-IMV rows. Figure~\ref{fig:SHAPIMV_Titanic} plots the same means with one sample standard deviation across seeds. Rankings describe predictive contributions within each specification, not causal effects.}
\label{SHAPIMVRankingtable}
\end{table*}

The full rankings in Table~\ref{SHAPIMVRankingtable} and seed variability in Figure~\ref{fig:SHAPIMV_Titanic} show where the estimators differ. For Titanic, sex ranks first under logistic regression (0.147), while title leads under both boosting models (0.113 each). All nevertheless select sex, title, class, age multiplied by class, and fare. Radius and concave points occupy the top two positions for Breast Cancer under every estimator. Married status leads for Adult Income, where the fifth selected predictor is weekly working hours for logistic regression and capital loss for boosting. These rankings describe predictive contributions within the specified models and feature sets, and do not imply causal effects.

\subsection{Multi-IMV Results}

\begin{figure*}[p]
    \centering
    \includegraphics[width=\linewidth]{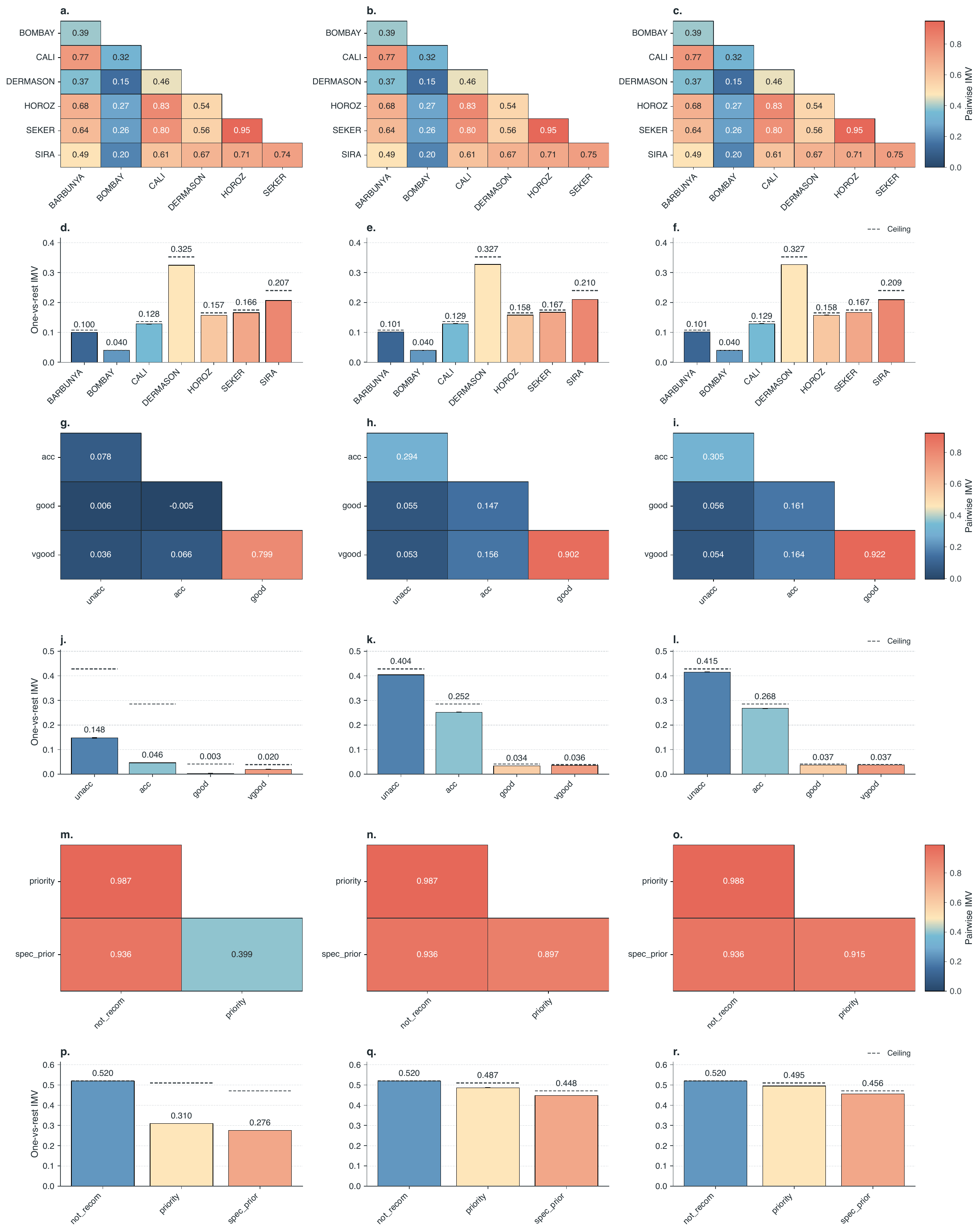}
    \caption{Multi-IMV on Dry Bean (\textbf{a.}--\textbf{f.}; seven classes, 16 features), Car Evaluation (\textbf{g.}--\textbf{l.}; four classes, six features), and Nursery (\textbf{m.}--\textbf{r.}; eight features, retaining not\_recom, priority, and spec\_prior). Columns show logistic regression, XGBoost, and LightGBM, each compared with its constant-input counterpart. Heatmaps restrict held-out observations to class pairs and renormalize both models' probabilities without binary refitting. Matrices are symmetric with conventional zero diagonals and one colour scale per dataset. Bars show one-vs-rest IMV from unrenormalized class probabilities on all held-out observations; the two views have different populations and reference coin weights. Five-fold stratified scores are averaged within each of ten seeds (42--51). Cells, bars, and labels show seed means; whiskers show one sample standard deviation, not confidence intervals. Dashed lines show the ceiling of the One-vs-rest IMV for the targeted category.}
    \label{fig:MultiIMV_Overview}
    \label{fig:MultiIMV_DryBean}
    \label{fig:MultiIMV_XGBoost}
    \label{fig:MultiIMV_CarNursery}
    \label{fig:MultiIMV_CarEvaluation}
    \label{fig:MultiIMV_Nursery}
    \label{fig:MultiIMV_LR}
    \label{fig:MultiIMV_LightGBM}
\end{figure*}

Table~\ref{tab:MultiIMVCarEvaluation} reports Multi-IMV, Brier loss, and one-vs-rest accuracy, precision, and recall for Car Evaluation. The reference for Multi-IMV is the same estimator fitted to a constant input. Entries are means and sample standard deviations across ten seeds, each evaluated with five stratified folds; global metrics summarize the four-class problem. 
For the rare \texttt{Good} class, logistic regression achieves 96.01\% one-vs-rest accuracy while never predicting the class, giving zero precision and recall. Its Brier loss is 0.04, yet mean Multi-IMV rounds to 0.00. High accuracy and low Brier loss can therefore give a misleading impression when read without the class frequencies or a reference model. XGBoost and LightGBM recover \texttt{Good} with recalls of 89.13\% and 95.22\%,  and their Multi-IMV scores of 0.03 and 0.04 are small in absolute terms, compared with 0.40 and 0.42 for \texttt{Unacceptable}. Relative to the bounds set by class frequency (0.04 for \texttt{Good}, 4.0\% of cars; 0.43 for \texttt{Unacceptable}, 70.0\%), however, both boosting models attain most of the available improvement for each class. Global support-weighted F1 is 63.84\%, 96.69\%, and 97.89\% for logistic regression, XGBoost, and LightGBM, respectively. 

Figure~\ref{fig:MultiIMV_Overview} places pairwise heatmaps alongside one-vs-rest bars for all three datasets and estimators. In Dry Bean, the pairwise IMV for Horoz/Seker exceeds that for Bombay/Dermason under every estimator. This contrast largely reflects class balance: Horoz and Seker have similar counts (bound 0.95), whereas Dermason outnumbers Bombay nearly seven to one
(bound 0.15), and every estimator comes within 0.003 of both bounds. For the
same reason, Dermason, the most frequent class, has the largest one-vs-rest
IMV but comparatively low pairwise scores. Dermason/Sira lies furthest below
its bound (0.67 of 0.74), identifying the pair the classifiers separate least
completely. These scores compare probabilistic fit between class pairs; they are not counts of classification errors and provide a different view from the one-vs-rest scores. Heatmaps use a common colour scale across estimators within each dataset. Further per-class metrics appear in Section~\ref{MoreresultsforMultiIMV}.

\begin{table*}[htbp]
\centering
\footnotesize
\begin{tabular}{lcccc}
\toprule
\multicolumn{1}{l}{{\bf{Methods}}} & \multicolumn{1}{c}{\bf{Unacc}} & \multicolumn{1}{c}{\bf{Acc}} & \multicolumn{1}{c}{\bf{Good}} & \multicolumn{1}{c}{\bf{Very good}}\\
\midrule
\textbf{Multi-IMV (LR)} & \textbf{0.15$\,\pm\,$0.00} & \textbf{0.05$\,\pm\,$0.00} & \textbf{0.00$\,\pm\,$0.00} & \textbf{0.02$\,\pm\,$0.00} \\
\rowcolor{gray!15}Brier Loss (LR) & 0.17$\,\pm\,$0.00 & 0.16$\,\pm\,$0.00 & 0.04$\,\pm\,$0.00 & 0.03$\,\pm\,$0.00 \\
Accuracy (LR) & 72.16$\,\pm\,$0.31 & 74.30$\,\pm\,$0.34 & 96.01$\,\pm\,$0.00 & 95.48$\,\pm\,$0.08 \\
\rowcolor{gray!15}Precision (LR) & 74.51$\,\pm\,$0.19 & 34.99$\,\pm\,$1.34 & 0.00$\,\pm\,$0.00 & 33.91$\,\pm\,$1.74 \\
Recall (LR) & 91.58$\,\pm\,$0.33 & 18.23$\,\pm\,$0.78 & 0.00$\,\pm\,$0.00 & 21.23$\,\pm\,$1.21 \\
\rowcolor{gray!15}\textbf{Multi-IMV (XGB)} & \textbf{0.40$\,\pm\,$0.00} & \textbf{0.25$\,\pm\,$0.00} & \textbf{0.03$\,\pm\,$0.00} & \textbf{0.04$\,\pm\,$0.00} \\
Brier Loss (XGB) & 0.02$\,\pm\,$0.00 & 0.03$\,\pm\,$0.00 & 0.01$\,\pm\,$0.00 & 0.01$\,\pm\,$0.00 \\
\rowcolor{gray!15}Accuracy (XGB) & 97.84$\,\pm\,$0.20 & 97.06$\,\pm\,$0.24 & 98.87$\,\pm\,$0.20 & 99.54$\,\pm\,$0.14 \\
Precision (XGB) & 99.32$\,\pm\,$0.20 & 91.93$\,\pm\,$0.89 & 83.67$\,\pm\,$1.77 & 91.85$\,\pm\,$2.50 \\
\rowcolor{gray!15}Recall (XGB) & 97.58$\,\pm\,$0.21 & 95.13$\,\pm\,$0.55 & 89.13$\,\pm\,$4.33 & 96.46$\,\pm\,$1.93 \\
\textbf{Multi-IMV (LGBM)} & \textbf{0.42$\,\pm\,$0.00} & \textbf{0.27$\,\pm\,$0.00} & \textbf{0.04$\,\pm\,$0.00} & \textbf{0.04$\,\pm\,$0.00} \\
\rowcolor{gray!15}Brier Loss (LGBM) & 0.01$\,\pm\,$0.00 & 0.02$\,\pm\,$0.00 & 0.01$\,\pm\,$0.00 & 0.00$\,\pm\,$0.00 \\
Accuracy (LGBM) & 98.67$\,\pm\,$0.20 & 98.13$\,\pm\,$0.14 & 99.18$\,\pm\,$0.19 & 99.73$\,\pm\,$0.10 \\
\rowcolor{gray!15}Precision (LGBM) & 99.84$\,\pm\,$0.17 & 94.67$\,\pm\,$0.64 & 85.90$\,\pm\,$2.14 & 95.10$\,\pm\,$1.66 \\
Recall (LGBM) & 98.26$\,\pm\,$0.21 & 97.06$\,\pm\,$0.61 & 95.22$\,\pm\,$3.13 & 98.00$\,\pm\,$1.63 \\
\bottomrule
\end{tabular}
\caption{One-vs-rest Multi-IMV and classification performance on Car Evaluation. LR, XGB, and LGBM denote logistic regression, XGBoost, and LightGBM. Values are means $\pm$ one sample standard deviation across 10 seeds, each using 5-fold stratified cross-validation; these are not confidence intervals. IMV compares the model using the predictors with the same estimator fitted to a constant input only. We average IMV over folds within each seed and calculate the other metrics from pooled out-of-fold predictions. Accuracy, precision, and recall are percentages for each class versus the rest, based on the multiclass argmax prediction; undefined precision or recall is set to zero. Brier Loss is the Brier score loss on class probabilities. Rounding to two decimals can produce a displayed zero standard deviation even when runs differ. Unacc and Acc denote Unacceptable and Acceptable. Global metrics (LR): F1 Score: 63.84$\,\pm\,$0.32; AUC-ROC: 77.29$\,\pm\,$0.12; Log Loss: 0.69$\,\pm\,$0.00; Hamming Loss: 0.31$\,\pm\,$0.00; Cohen's Kappa: 16.31$\,\pm\,$0.78; Jaccard: 52.42$\,\pm\,$0.31.
Global metrics (XGB): F1 Score: 96.69$\,\pm\,$0.29; AUC-ROC: 99.71$\,\pm\,$0.05; Log Loss: 0.14$\,\pm\,$0.00; Hamming Loss: 0.03$\,\pm\,$0.00; Cohen's Kappa: 92.79$\,\pm\,$0.63; Jaccard: 93.76$\,\pm\,$0.51.
Global metrics (LGBM): F1 Score: 97.89$\,\pm\,$0.18; AUC-ROC: 99.90$\,\pm\,$0.04; Log Loss: 0.08$\,\pm\,$0.00; Hamming Loss: 0.02$\,\pm\,$0.00; Cohen's Kappa: 95.38$\,\pm\,$0.40; Jaccard: 95.94$\,\pm\,$0.34.
F1, one-vs-rest AUC-ROC, and Jaccard use support-weighted class averages. These three metrics and Cohen's kappa are multiplied by 100; log loss and Hamming loss are unscaled.}
\label{tab:MultiIMVCarEvaluation}
\end{table*}

\subsubsection{Nursery and Dry Bean Comparisons}\label{MoreresultsforMultiIMV}
The Dry Bean, Car Evaluation, and Nursery results occupy panels \textbf{a.}--\textbf{f.}, \textbf{g.}--\textbf{l.}, and \textbf{m.}--\textbf{r.} of Figure~\ref{fig:MultiIMV_Overview}, respectively. Each dataset has a row of pairwise heatmaps and a row of one-vs-rest bars for logistic regression, XGBoost, and LightGBM. All four Car Evaluation classes are retained. For Nursery, the analysis retains \texttt{not\_recom}, \texttt{priority}, and \texttt{spec\_prior}, excluding \texttt{recommend} and \texttt{very\_recom}. Tables~\ref{tab:MUltiIMVNursery} and \ref{tab:MUltiIMVDryBean} report the additional per-class metrics for Nursery and Dry Bean.

\begin{table*}[htbp]
\centering
\footnotesize
\begin{tabular}{lccc}
\toprule
\multicolumn{1}{l}{{\bf{Methods}}} & \multicolumn{1}{c} {\bf{Not recommend}} & \multicolumn{1}{c} {\bf{Priority acceptance}} & \multicolumn{1}{c} {\bf{Special acceptance}} \\
\midrule
\textbf{Multi-IMV (LR)} & \textbf{0.52$\,\pm\,$0.00} & \textbf{0.31$\,\pm\,$0.00} & \textbf{0.28$\,\pm\,$0.00} \\
\rowcolor{gray!15}Brier Loss (LR) & 0.00$\,\pm\,$0.00 & 0.13$\,\pm\,$0.00 & 0.13$\,\pm\,$0.00 \\
Accuracy (LR) & 100.00$\,\pm\,$0.00 & 78.75$\,\pm\,$0.05 & 78.75$\,\pm\,$0.05 \\
\rowcolor{gray!15}Precision (LR) & 100.00$\,\pm\,$0.00 & 68.20$\,\pm\,$0.07 & 67.16$\,\pm\,$0.10 \\
Recall (LR) & 100.00$\,\pm\,$0.00 & 69.48$\,\pm\,$0.12 & 65.82$\,\pm\,$0.10 \\
\rowcolor{gray!15}\textbf{Multi-IMV (XGB)} & \textbf{0.52$\,\pm\,$0.00} & \textbf{0.49$\,\pm\,$0.00} & \textbf{0.45$\,\pm\,$0.00}\\
Brier Loss (XGB) & 0.00$\,\pm\,$0.00 & 0.02$\,\pm\,$0.00 & 0.02$\,\pm\,$0.00 \\
\rowcolor{gray!15}Accuracy (XGB) & 100.00$\,\pm\,$0.00 & 99.40$\,\pm\,$0.05 & 99.40$\,\pm\,$0.05 \\
Precision (XGB) & 100.00$\,\pm\,$0.00 & 99.92$\,\pm\,$0.03 & 98.24$\,\pm\,$0.16 \\
\rowcolor{gray!15}Recall (XGB) & 100.00$\,\pm\,$0.00 & 98.31$\,\pm\,$0.15 & 99.92$\,\pm\,$0.03 \\
\textbf{Multi-IMV (LGBM)} & \textbf{0.52$\,\pm\,$0.00} & \textbf{0.49$\,\pm\,$0.00} & \textbf{0.46$\,\pm\,$0.00} \\
\rowcolor{gray!15}Brier Loss (LGBM) & 0.00$\,\pm\,$0.00 & 0.01$\,\pm\,$0.00 & 0.01$\,\pm\,$0.00 \\
Accuracy (LGBM) & 100.00$\,\pm\,$0.00 & 99.76$\,\pm\,$0.06 & 99.76$\,\pm\,$0.06 \\
\rowcolor{gray!15}Precision (LGBM) & 100.00$\,\pm\,$0.00 & 99.98$\,\pm\,$0.03 & 99.29$\,\pm\,$0.17 \\
Recall (LGBM) & 100.00$\,\pm\,$0.00 & 99.32$\,\pm\,$0.17 & 99.98$\,\pm\,$0.03 \\
\bottomrule
\end{tabular}
\caption{Multi-IMV results for Nursery using logistic regression, XGBoost, and LightGBM. Means and sample standard deviations are calculated across ten seeds, each evaluated with five stratified folds. Metric definitions and units follow Table~\ref{tab:MultiIMVCarEvaluation}. Global metrics (LR): F1 Score: 78.74$\,\pm\,$0.05; AUC-ROC: 91.77$\,\pm\,$0.01; Log Loss: 0.39$\,\pm\,$0.00; Hamming Loss: 0.21$\,\pm\,$0.00; Cohen's Kappa: 68.11$\,\pm\,$0.08; Jaccard: 67.87$\,\pm\,$0.06. 
Global metrics (XGB): F1 Score: 99.40$\,\pm\,$0.05; AUC-ROC: 99.99$\,\pm\,$0.00; Log Loss: 0.08$\,\pm\,$0.00; Hamming Loss: 0.01$\,\pm\,$0.00; Cohen's Kappa: 99.10$\,\pm\,$0.08; Jaccard: 98.82$\,\pm\,$0.10. 
Global metrics (LGBM): F1 Score: 99.76$\,\pm\,$0.06; AUC-ROC: 100.00$\,\pm\,$0.00; Log Loss: 0.06$\,\pm\,$0.00; Hamming Loss: 0.00$\,\pm\,$0.00; Cohen's Kappa: 99.65$\,\pm\,$0.09; Jaccard: 99.53$\,\pm\,$0.12.} 
\label{tab:MUltiIMVNursery}
\end{table*}

\begin{table*}[htbp]
\centering
\footnotesize
\begin{tabular}{lccccccc}
\toprule
\multicolumn{1}{l}{{\bf{Methods}}} & \multicolumn{1}{c} {\bf{Seker}} & \multicolumn{1}{c} {\bf{Barbunya}} & \multicolumn{1}{c} {\bf{Bombay}} & \multicolumn{1}{c}{\bf{Cali}} & \multicolumn{1}{c}{\bf{Dermason}} & \multicolumn{1}{c}{\bf{Horoz}} & \multicolumn{1}{c}{\bf{Sira}}\\
\midrule
\textbf{Multi-IMV (LR)} & \textbf{0.17$\,\pm\,$0.00} & \textbf{0.10$\,\pm\,$0.00} & \textbf{0.04$\,\pm\,$0.00} & \textbf{0.13$\,\pm\,$0.00} & \textbf{0.32$\,\pm\,$0.00} & \textbf{0.16$\,\pm\,$0.00} & \textbf{0.21$\,\pm\,$0.00} \\
\rowcolor{gray!15}Brier Loss (LR) & 0.01$\,\pm\,$0.00 & 0.01$\,\pm\,$0.00 & 0.00$\,\pm\,$0.00 & 0.01$\,\pm\,$0.00 & 0.03$\,\pm\,$0.00 & 0.01$\,\pm\,$0.00 & 0.04$\,\pm\,$0.00\\
Accuracy (LR) & 98.45$\,\pm\,$0.02 & 98.55$\,\pm\,$0.03 & 100.00$\,\pm\,$0.00 & 98.53$\,\pm\,$0.01 & 95.83$\,\pm\,$0.03 & 98.71$\,\pm\,$0.01 & 94.76$\,\pm\,$0.04\\
\rowcolor{gray!15}Precision (LR) & 94.93$\,\pm\,$0.10 & 94.22$\,\pm\,$0.13 & 100.00$\,\pm\,$0.00 & 93.41$\,\pm\,$0.10 & 92.28$\,\pm\,$0.06 & 95.47$\,\pm\,$0.10 & 85.61$\,\pm\,$0.12\\
Recall (LR) & 94.68$\,\pm\,$0.08 & 90.63$\,\pm\,$0.22 & 99.96$\,\pm\,$0.08 & 94.36$\,\pm\,$0.06 & 91.66$\,\pm\,$0.06 & 95.41$\,\pm\,$0.06 & 87.69$\,\pm\,$0.12\\
\rowcolor{gray!15}\textbf{Multi-IMV (XGB)} & \textbf{0.17$\,\pm\,$0.00} & \textbf{0.10$\,\pm\,$0.00} & \textbf{0.04$\,\pm\,$0.00} & \textbf{0.13$\,\pm\,$0.00} & \textbf{0.33$\,\pm\,$0.00} & \textbf{0.16$\,\pm\,$0.00} & \textbf{0.21$\,\pm\,$0.00}\\
Brier Loss (XGB) & 0.01$\,\pm\,$0.00 & 0.01$\,\pm\,$0.00 & 0.00$\,\pm\,$0.00 & 0.01$\,\pm\,$0.00 & 0.03$\,\pm\,$0.00 & 0.01$\,\pm\,$0.00 & 0.03$\,\pm\,$0.00\\
\rowcolor{gray!15}Accuracy (XGB) & 98.56$\,\pm\,$0.03 & 98.73$\,\pm\,$0.06 & 99.98$\,\pm\,$0.01 & 98.65$\,\pm\,$0.06 & 95.78$\,\pm\,$0.08 & 98.74$\,\pm\,$0.03 & 95.07$\,\pm\,$0.07\\
Precision (XGB) & 95.50$\,\pm\,$0.14 & 94.60$\,\pm\,$0.50 & 99.62$\,\pm\,$0.09 & 94.12$\,\pm\,$0.31 & 90.99$\,\pm\,$0.16 & 95.83$\,\pm\,$0.10 & 87.67$\,\pm\,$0.31\\
\rowcolor{gray!15}Recall (XGB) & 94.81$\,\pm\,$0.24 & 92.20$\,\pm\,$0.27 & 99.89$\,\pm\,$0.13 & 94.61$\,\pm\,$0.29 & 92.99$\,\pm\,$0.23 & 95.28$\,\pm\,$0.19 & 86.71$\,\pm\,$0.15\\
\textbf{Multi-IMV (LGBM)} & \textbf{0.17$\,\pm\,$0.00} & \textbf{0.10$\,\pm\,$0.00} & \textbf{0.04$\,\pm\,$0.00} & \textbf{0.13$\,\pm\,$0.00} & \textbf{0.33$\,\pm\,$0.00} & \textbf{0.16$\,\pm\,$0.00} & \textbf{0.21$\,\pm\,$0.00}\\
\rowcolor{gray!15}Brier Loss (LGBM) & 0.01$\,\pm\,$0.00 & 0.01$\,\pm\,$0.00 & 0.00$\,\pm\,$0.00 & 0.01$\,\pm\,$0.00 & 0.03$\,\pm\,$0.00 & 0.01$\,\pm\,$0.00 & 0.04$\,\pm\,$0.00\\
Accuracy (LGBM) & 98.55$\,\pm\,$0.03 & 98.75$\,\pm\,$0.04 & 99.98$\,\pm\,$0.00 & 98.67$\,\pm\,$0.06 & 95.76$\,\pm\,$0.07 & 98.76$\,\pm\,$0.03 & 95.05$\,\pm\,$0.09\\
\rowcolor{gray!15}Precision (LGBM) & 95.35$\,\pm\,$0.21 & 94.60$\,\pm\,$0.21 & 99.73$\,\pm\,$0.10 & 94.36$\,\pm\,$0.31 & 91.02$\,\pm\,$0.19 & 95.92$\,\pm\,$0.28 & 87.52$\,\pm\,$0.31\\
Recall (LGBM) & 94.87$\,\pm\,$0.22 & 92.38$\,\pm\,$0.34 & 99.77$\,\pm\,$0.08 & 94.53$\,\pm\,$0.26 & 92.90$\,\pm\,$0.16 & 95.27$\,\pm\,$0.10 & 86.79$\,\pm\,$0.23\\
\bottomrule
\end{tabular}
\caption{Multi-IMV results for Dry Bean using logistic regression, XGBoost, and LightGBM. Means and sample standard deviations are calculated across ten seeds, each evaluated with five stratified folds. Metric definitions and units follow Table~\ref{tab:MultiIMVCarEvaluation}. Global metrics (LR): F1 Score: 92.43$\,\pm\,$0.04; AUC-ROC: 99.34$\,\pm\,$0.00; Log Loss: 0.21$\,\pm\,$0.00; Hamming Loss: 0.08$\,\pm\,$0.00; Cohen's Kappa: 90.83$\,\pm\,$0.05; Jaccard: 86.10$\,\pm\,$0.06. 
Global metrics (XGB): F1 Score: 92.75$\,\pm\,$0.11; AUC-ROC: 99.42$\,\pm\,$0.01; Log Loss: 0.19$\,\pm\,$0.00; Hamming Loss: 0.07$\,\pm\,$0.00; Cohen's Kappa: 91.23$\,\pm\,$0.14; Jaccard: 86.65$\,\pm\,$0.19.
Global metrics (LGBM): F1 Score: 92.75$\,\pm\,$0.11; AUC-ROC: 99.41$\,\pm\,$0.01; Log Loss: 0.20$\,\pm\,$0.00; Hamming Loss: 0.07$\,\pm\,$0.00; Cohen's Kappa: 91.23$\,\pm\,$0.13; Jaccard: 86.66$\,\pm\,$0.18.}
\label{tab:MUltiIMVDryBean}
\end{table*}

\subsection{Ablate-IMV Results}

\begin{figure*}[htbp]
    \centering
    \includegraphics[width=\linewidth]{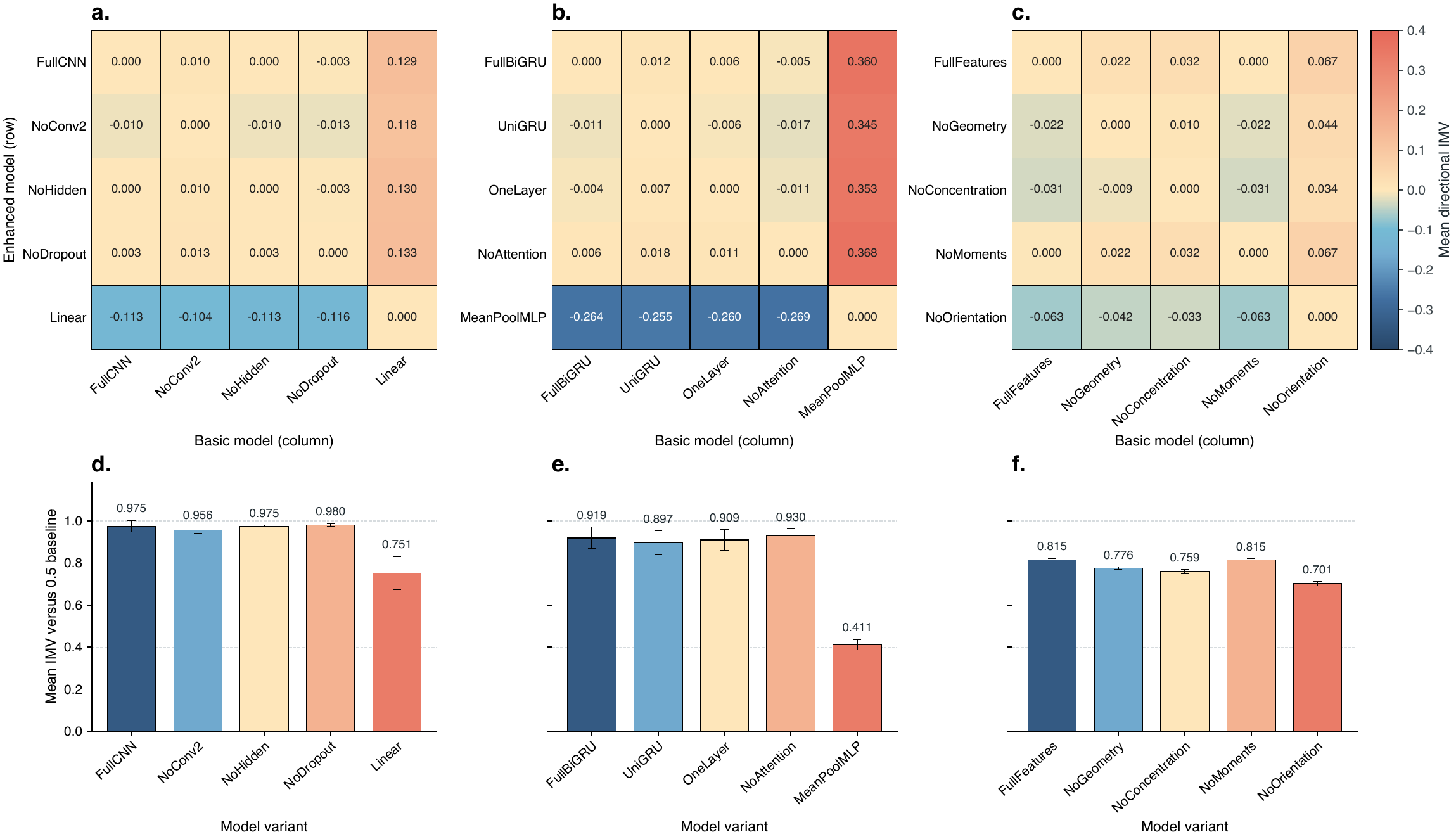}
    \caption{Ablate-IMV for MNIST odd/even digits (CNN; \textbf{a.}, \textbf{d.}), UCI HAR upstairs/downstairs activity (bidirectional GRU; \textbf{b.}, \textbf{e.}), and MAGIC Gamma gamma/hadron events (MLP; \textbf{c.}, \textbf{f.}). Heatmaps compare enhanced rows with basic columns; positive values favor rows, and reverse comparisons need not be negatives. Diagonals are zero; heatmaps share a zero-centred colour scale. Bars compare models with constant 0.5 predictions, not the full model shown first. Cells, bars, and labels show ten-seed means (42--51); whiskers show one sample standard deviation, not confidence intervals. MNIST/HAR ablate architecture; MAGIC removes feature groups with hidden architecture fixed. Training lasts one, 15, and 12 epochs, respectively. Official MNIST/HAR test sets and stratified 75/25 MAGIC splits align observations across variants within seeds.}
    \label{fig:Ablate-IMV-Overview}
    \label{fig:Ablate-IMV-NLP}
    \label{fig:Ablate-IMV-CV}
\end{figure*}

Figure~\ref{fig:Ablate-IMV-Overview} compares architectural ablations on MNIST and UCI HAR and feature-group ablations on MAGIC Gamma. In each heatmap, the row is the enhanced model and the column is its reference; the bars instead compare every model with constant 0.5-probability predictions. Variants are evaluated on the same test observations within each seed. Means are taken over ten runs (42--51), and whiskers describe variation across seeds rather than confidence intervals.

\smallskip\subsubsection{MNIST: Architectural Ablation of a CNN}
The CNN in panels \textbf{a.} and \textbf{d.} classifies odd/even digits using two convolutions, max pooling, dropout, and a hidden layer. Variants remove the second convolution (\texttt{NoConv2}), hidden layer (\texttt{NoHidden}), or dropout (\texttt{NoDropout}), or replace the convolutional and hidden stack with a linear classifier (\texttt{Linear}). After one epoch, the linear model has a mean IMV of 0.751 against the 0.5 baseline, compared with 0.975 for the full CNN. The other ablations are closer to the full model; removing dropout raises the score slightly to 0.980. Under this training budget, removing a component need not reduce held-out fit.

\smallskip\subsubsection{UCI HAR: Recurrent and Temporal Components}
Panels \textbf{b.} and \textbf{e.} distinguish upstairs/downstairs activity from smartphone inertial sequences, using the official participant-disjoint partition. The full model has two bidirectional GRU layers and temporal attention. Variants use one recurrent direction (\texttt{UniGRU}), one layer (\texttt{OneLayer}), mean pooling instead of attention (\texttt{NoAttention}), or an MLP using only channel means (\texttt{MeanPoolMLP}). The mean-only MLP scores 0.411 against the 0.5 baseline, compared with 0.919 for the full model. Mean pooling over recurrent states scores slightly higher than the full model, at 0.930. The full recurrent model has higher mean fit than the channel-mean MLP, but its attention mechanism gives no gain over recurrent mean pooling under this training budget.

\smallskip\subsubsection{MAGIC Gamma: Feature-Group Ablation}
Panels \textbf{c.} and \textbf{f.} classify gamma/hadron events with a fixed MLP hidden architecture. Ablations remove geometry (\texttt{NoGeometry}), size and concentration (\texttt{NoConcentration}), moments (\texttt{NoM} \texttt{oments}), or orientation and distance (\texttt{NoOrientation}). All variants share seed-specific stratified 75/25 splits and standardization fitted on the training data. Removing orientation and distance gives the largest decrease in mean IMV against the 0.5 baseline (0.701 versus 0.815 for all features); removing moments leaves the mean almost unchanged (0.815). The heatmap allows direct comparisons between variants. The apparent usefulness of these feature groups remains conditional on the specified architecture, training budget, and dataset.
\section{IMV and Model Training: A Case Study}

\begin{figure*}[htbp]
    \centering
    \includegraphics[width=\linewidth]{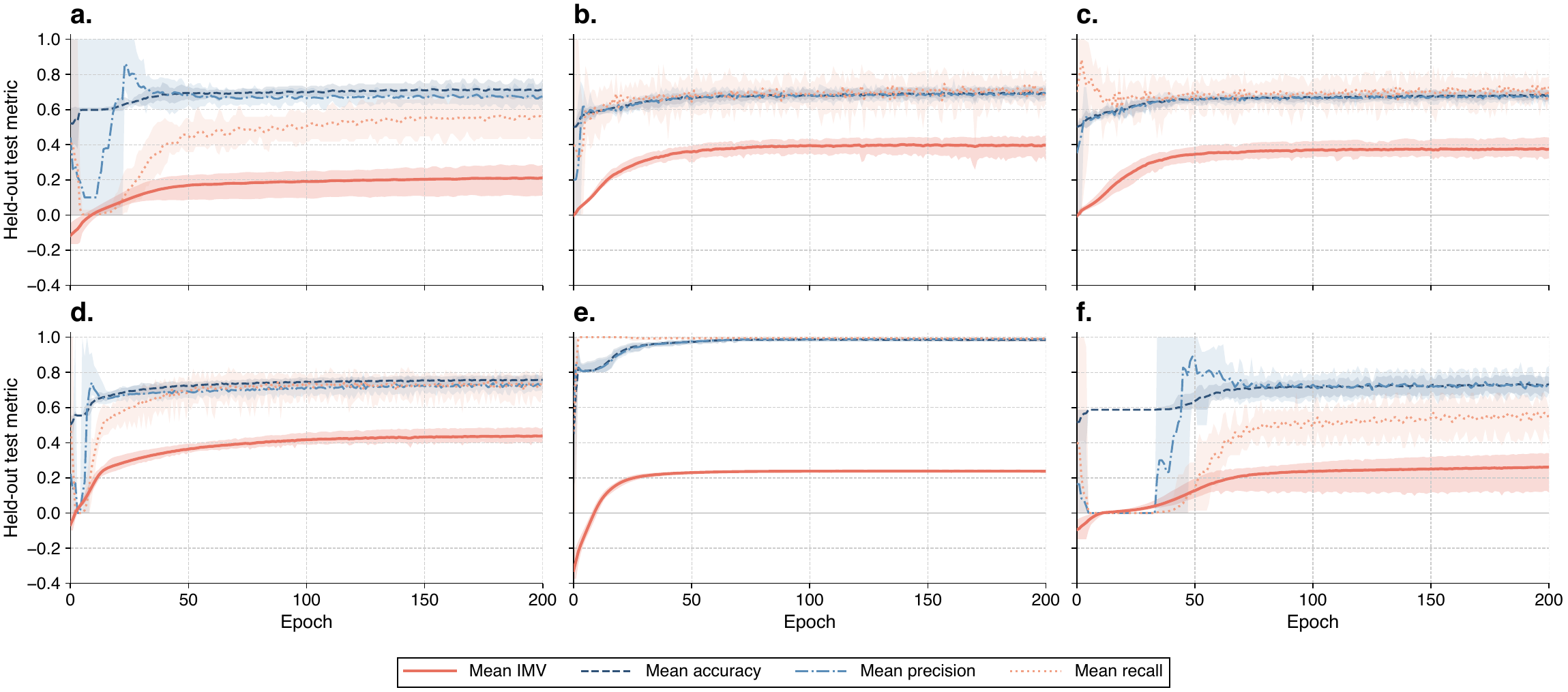}
    \caption{Test performance during training of a three-layer GCN on \textbf{a.} PROTEINS (1,113 graphs), \textbf{b.} NCI1 (4,110), \textbf{c.} NCI109 (4,127), \textbf{d.} Mutagenicity (4,337), \textbf{e.} AIDS (2,000), and \textbf{f.} DD (1,178). Counts refer to full datasets. Lines show means across ten seeds (42--51): solid for IMV, dashed for accuracy, dash-dotted for precision, and dotted for recall. Shading spans the minimum and maximum across seeds at each epoch, not confidence intervals. Each seed defines a stratified 70/15/15 training/validation/test split; the same test graphs are evaluated from untrained epoch 0 through epoch 200. IMV compares probabilities with the training-prevalence null; classification uses a 0.5 threshold with class 1 positive. Curves are not smoothed, and negative IMV values are retained.}
    \label{fig:GNN-Training}
    \label{fig:combindIMVepoch}
\end{figure*}

\subsection{Graph Datasets and Preprocessing}\label{datasets:gnn}
We evaluate six binary graph-classification benchmarks drawn from TUDataset~\cite{morris2020tudataset}; Table~\ref{tab:gnn-datasets} reports graph counts and sizes. PROTEINS links secondary-structure elements; DD links nearby amino acids in larger graphs. Both distinguish enzymes from non-enzymes~\cite{Protein,shervashidze2011weisfeiler}. NCI1 and NCI109 contain molecules screened for activity against non-small-cell lung and ovarian cancer cell lines, respectively~\cite{NCI1NCI109,shervashidze2011weisfeiler}. Mutagenicity comprises 2{,}401 Ames-test mutagens and 1{,}936 nonmutagens~\cite{kazius2005mutagenicity}; AIDS comprises 400 HIV-active and 1{,}600 inactive compounds~\cite{riesen2008iam}. Molecular graphs represent atoms as nodes and bonds as edges. Following TU label ordering, class one denotes nonmutagenicity in Mutagenicity and inactivity in AIDS.

\begin{table}[htbp]
\centering
\footnotesize
\setlength{\tabcolsep}{4pt}
\begin{tabular}{lrrr}
\toprule
Dataset & Graphs & Mean nodes & Mean edges \\
\midrule
PROTEINS & 1{,}113 & 39.06 & 72.82 \\
NCI1 & 4{,}110 & 29.87 & 32.30 \\
NCI109 & 4{,}127 & 29.68 & 32.13 \\
Mutagenicity & 4{,}337 & 30.32 & 30.77 \\
AIDS & 2{,}000 & 15.69 & 16.20 \\
DD & 1{,}178 & 284.32 & 715.66 \\
\bottomrule
\end{tabular}
\caption{Datasets used in the GNN experiments. Mean edge counts refer to undirected edges before GCN self-loops are added.}
\label{tab:gnn-datasets}
\end{table}

PyTorch Geometric loads the original TU releases, retaining all graphs and categorical node labels~\cite{Fey/Lenssen/2019}. Numeric node attributes (PROTEINS: one; AIDS: four) are standardized using training nodes only; bond labels are unused. Each seed defines an approximately 70\%/15\%/15\% stratified random training/validation/test split, without isomorphism cleaning or molecular scaffold-based splitting.

\subsection{Architecture and Training}
The GCN comprises three graph-convolution layers with symmetric normalization, self-loops, and ReLU, followed by global mean pooling, dropout, and a linear binary-logit output. Adam minimizes binary cross-entropy without weight decay, learning-rate scheduling, or early stopping. Each optimizer step covers up to 512 graphs through node-count-limited microbatches with sample-weighted gradient accumulation. Validation metrics are recorded but never used for tuning or checkpoint selection; test curves follow the full fixed training run.

\subsection{Epoch-Level Results}
Figure~\ref{fig:GNN-Training} tracks each seed's fixed test graphs at every epoch from untrained epoch 0 through 200, using ten seeds (42--51) per dataset. Lines show seed means; shading spans the epoch-wise minimum and maximum. These ranges reflect variation across splits and training runs, not confidence intervals. All epochs and negative IMV values are retained, without smoothing or path simplification; a steady mean can mask variation between runs. IMV compares probabilities with the training-prevalence null. Accuracy, binary precision, and binary recall use threshold 0.5 with class 1 positive, so IMV can vary without changes in predicted labels or these metrics. Section~\ref{Complexity} reports computational costs.

\section{Computational Complexity Analysis}\label{Complexity}

For the \texttt{vanilla\_imv} call in Box~\ref{cb:vanilla-imv}, validating $n$ test observations and computing two geometric mean Bernoulli likelihoods take $\mathcal{O}(n)$ time. The default Brent root finder then solves for each likelihood's equivalent coin weight. If the two scalar searches require $k_0$ and $k_1$ iterations, total time is $\mathcal{O}(n+k_0+k_1)$. At fixed numerical precision, these searches do not operate on the observation vectors, so computation scales linearly with $n$. The current vectorised implementation requires $\mathcal{O}(n)$ auxiliary memory for the broadcast baseline and temporary arrays. 
We extend the simulation in Box~\ref{cb:vanilla-imv} to total sample sizes of 1,000, 10,000, 100,000, and 1,000,000. The data-generating process, random seed, logistic regression, and 70/30 training/test split stay the same, with training-set prevalence as the baseline. Timing begins after data generation, model fitting, and test prediction. It covers the complete call \texttt{vanilla\_imv(p\_baseline, p\_enhanced, y\_test)} with default settings, including input validation, baseline broadcasting, likelihood computation, and root finding. Runs are sequential and use one numerical-library thread. 
After a warm-up call, Python's \texttt{timeit} calibrates a batch lasting at least 0.2 seconds. We run nine batches of that size on identical inputs, dividing durations by call counts. Table~\ref{tab:time} reports the median and interquartile range of per-call times and the median time per test observation. The interquartile range describes timing variation, not IMV uncertainty. The notebook \texttt{src/simulations/complexity.ipynb} generates the table and saves individual batches and software and hardware metadata. Fixed overhead, including the two scalar root searches, raises the cost per observation at small sample sizes. At larger sample sizes, the cost per observation varies little, consistent with the linear dependence on $n$. These timings cover IMV calculation alone, excluding fitting and prediction, and are not total training times per epoch.

\begin{table}[htbp]
\centering
\footnotesize
\setlength{\tabcolsep}{4pt}
\begin{tabular}{rrrrr}
\toprule
$n$ & IMV & \shortstack{Median\\(ms)} & \shortstack{IQR\\(ms)} & \shortstack{Time/obs.\\($\mu$s)} \\
\midrule
300 & 0.282 & 0.078 & 0.002 & 0.260 \\
3,000 & 0.279 & 0.124 & 0.000 & 0.041 \\
30,000 & 0.303 & 0.719 & 0.010 & 0.024 \\
300,000 & 0.303 & 6.987 & 0.319 & 0.023 \\
\bottomrule
\end{tabular}
\caption{Runtime of \texttt{vanilla\_imv} for the simulation in Box~\ref{cb:vanilla-imv}. The median and interquartile range (IQR) summarize 9 batches of repeated calls; $n$ is the number of test observations. Timing excludes fitting and prediction.}
\label{tab:time}
\par\smallskip
\raggedright\footnotesize
\texttt{imvpy} 1.2.0; Python 3.12.7; Intel(R) Core(TM) i9-14900KF; one numerical-library thread.
\end{table}

\FloatBarrier
\section{Conclusion}
\label{sec::conclusion}

SHAP-IMV selected predictor subsets that performed comparably to those chosen by SHAP, LIME, and Anchors, while expressing contributions as held-out gains in fit; its rankings changed with the estimator. Multi-IMV gains differed across classes, and Ablate-IMV showed that removing model components need not reduce fit under our training budgets. The GNN results show how IMV and threshold-based metrics can change differently during training. These findings depend on the baseline, evaluation design, and coin-mapping domain; they neither guarantee better prediction nor establish causal effects. Exact attribution requires substantial refitting. The \texttt{imvpy} package~\cite{imvpy2026} implements these three comparisons of probabilistic fit against an explicit baseline.

\paragraph{Code and data availability.} The \href{https://github.com/intermodelvigorish/IMV_ML}{IMV\_ML repository} provides the notebooks and plotting code under GNU GPL v3. The notebooks retrieve the cited datasets, generate local results, and display validation summaries inline; \texttt{run\_all.sh} runs the pipeline. \texttt{imvpy} is published on PyPI~\cite{imvpy2026}, and the analysis pins a Git commit.

\renewcommand{\bibfont}{\RaggedRight}
\setlength{\bibsep}{1pt plus 0.5pt minus 0.5pt}
\bibliographystyle{unsrtnat}
\bibliography{references}

@article{zhangintermodel,
  title={Evaluating Model Predictive Performance in Confirmatory Factor Analysis with Binary Outcomes Using the {InterModel Vigorish}},
  author={Zhang, Lijin and Rahal, Charles and Kanopka, Klint and Ulitzsch, Esther and Zhang, Zhiyong and Domingue, Benjamin W.},
  journal={Multivariate Behavioral Research},
  pages={1--20},
  year={2026},
  doi={10.1080/00273171.2026.2645212},
  note={Advance online publication}
}

@article{domingue2021intermodel,
  title={The InterModel Vigorish (IMV) as a flexible and portable approach for quantifying predictive accuracy with binary outcomes},
  author={Domingue, Benjamin W. and Rahal, Charles and Faul, Jessica and Freese, Jeremy and Kanopka, Klint and Rigos, Alexandros and Stenhaug, Ben and Tripathi, Ajay Shanker},
  journal={PLOS ONE},
  volume={20},
  number={3},
  pages={e0316491},
  year={2025},
  doi={10.1371/journal.pone.0316491}
}

@article{domingue2024intermodel,
  title={The InterModel Vigorish as a lens for understanding (and quantifying) the value of item response models for dichotomously coded items},
  author={Domingue, Benjamin W and Kanopka, Klint and Kapoor, Radhika and Pohl, Steffi and Chalmers, R Philip and Rahal, Charles and Rhemtulla, Mijke},
  journal={Psychometrika},
  volume={89},
  number={3},
  pages={1034--1054},
  year={2024},
  doi={10.1007/s11336-024-09977-2}
}

@inproceedings{SHAP,
  title={A unified approach to interpreting model predictions},
  author={Lundberg, Scott M. and Lee, Su-In},
  booktitle={Advances in Neural Information Processing Systems},
  volume={30},
  pages={4765--4774},
  year={2017},
  publisher={Curran Associates, Inc}
}

@inproceedings{LIME,
  title={" Why should i trust you?" Explaining the predictions of any classifier},
  author={Ribeiro, Marco Tulio and Singh, Sameer and Guestrin, Carlos},
  booktitle={Proceedings of the 22nd ACM SIGKDD international conference on knowledge discovery and data mining},
  pages={1135--1144},
  year={2016}
}

@inproceedings{Anchors,
  title={Anchors: High-precision model-agnostic explanations},
  author={Ribeiro, Marco Tulio and Singh, Sameer and Guestrin, Carlos},
  booktitle={Proceedings of the AAAI conference on artificial intelligence},
  volume={32},
  year={2018},
  doi={10.1609/aaai.v32i1.11491}
}

@article{Entropy,
  title={A mathematical theory of communication},
  author={Shannon, Claude Elwood},
  journal={The Bell system technical journal},
  volume={27},
  number={3},
  pages={379--423},
  year={1948},
  publisher={Nokia Bell Labs}
}

@article{decisiontrees,
  title={Induction of decision trees},
  author={Quinlan, J. Ross},
  journal={Machine Learning},
  volume={1},
  pages={81--106},
  year={1986},
  publisher={Springer}
}

@article{Randomforests,
  title={Random forests},
  author={Breiman, Leo},
  journal={Machine learning},
  volume={45},
  pages={5--32},
  year={2001},
  publisher={Springer}
}

@inproceedings{gordon2021disagreement,
  title={The disagreement deconvolution: Bringing machine learning performance metrics in line with reality},
  author={Gordon, Mitchell L and Zhou, Kaitlyn and Patel, Kayur and Hashimoto, Tatsunori and Bernstein, Michael S},
  booktitle={Proceedings of the 2021 CHI Conference on Human Factors in Computing Systems},
  pages={1--14},
  year={2021}
}

@article{hase2020evaluating,
  title={Evaluating explainable AI: Which algorithmic explanations help users predict model behavior?},
  author={Hase, Peter and Bansal, Mohit},
  journal={arXiv preprint arXiv:2005.01831},
  year={2020}
}

@article{R^2,
  title={VII. Mathematical contributions to the theory of evolution.—III. Regression, heredity, and panmixia},
  author={Pearson, Karl},
  journal={Philosophical Transactions of the Royal Society of London. Series A, containing papers of a mathematical or physical character},
  volume={187},
  pages={253--318},
  year={1896},
  publisher={The Royal Society London}
}

@book{Accuracy,
  title={Discriminatory analysis: nonparametric discrimination, consistency properties},
  author={Fix, Evelyn},
  volume={1},
  year={1985},
  publisher={USAF school of Aviation Medicine}
}

@article{flach2015precision,
  title={Precision-recall-gain curves: PR analysis done right},
  author={Flach, Peter and Kull, Meelis},
  journal={Advances in neural information processing systems},
  volume={28},
  year={2015}
}

@article{hanley1982meaning,
  title={The meaning and use of the area under a receiver operating characteristic (ROC) curve.},
  author={Hanley, James A and McNeil, Barbara J},
  journal={Radiology},
  volume={143},
  number={1},
  pages={29--36},
  year={1982}
}

@article{AIC,
  title={A new look at the statistical model identification},
  author={Akaike, Hirotugu},
  journal={IEEE transactions on automatic control},
  volume={19},
  number={6},
  pages={716--723},
  year={1974},
  publisher={Ieee}
}

@article{BIC,
  title={Estimating the dimension of a model},
  author={Schwarz, Gideon},
  journal={The annals of statistics},
  pages={461--464},
  year={1978},
  publisher={JSTOR}
}

@article{shapley1953value,
  title={A value for n-person games},
  author={Shapley, Lloyd S},
  journal={Contribution to the Theory of Games},
  volume={2},
  year={1953}
}

@article{paszke2019pytorch,
  title={Pytorch: An imperative style, high-performance deep learning library},
  author={Paszke, Adam and Gross, Sam and Massa, Francisco and Lerer, Adam and Bradbury, James and Chanan, Gregory and Killeen, Trevor and Lin, Zeming and Gimelshein, Natalia and Antiga, Luca and others},
  journal={Advances in neural information processing systems},
  volume={32},
  year={2019}
}

@article{pedregosa2011scikit,
  title={Scikit-learn: Machine learning in Python},
  author={Pedregosa, Fabian and Varoquaux, Ga{\"e}l and Gramfort, Alexandre and Michel, Vincent and Thirion, Bertrand and Grisel, Olivier and Blondel, Mathieu and Prettenhofer, Peter and Weiss, Ron and Dubourg, Vincent and others},
  journal={the Journal of machine Learning research},
  volume={12},
  pages={2825--2830},
  year={2011},
  publisher={JMLR. org}
}

@inproceedings{chen2016xgboost,
  title={Xgboost: A scalable tree boosting system},
  author={Chen, Tianqi and Guestrin, Carlos},
  booktitle={Proceedings of the 22nd acm sigkdd international conference on knowledge discovery and data mining},
  pages={785--794},
  year={2016}
}

@article{ke2017lightgbm,
  title={Lightgbm: A highly efficient gradient boosting decision tree},
  author={Ke, Guolin and Meng, Qi and Finley, Thomas and Wang, Taifeng and Chen, Wei and Ma, Weidong and Ye, Qiwei and Liu, Tie-Yan},
  journal={Advances in neural information processing systems},
  volume={30},
  year={2017}
}

@misc{asuncion2007uci,
  title={{UCI Machine Learning Repository}},
  author={Asuncion, Arthur and Newman, David and others},
  year={2007},
  publisher={Irvine, CA, USA}
}

@misc{adult_2,
  author       = {Becker, Barry and Kohavi, Ronny},
  title        = {{Adult}},
  year         = {1996},
  howpublished = {{UCI Machine Learning Repository}},
  note         = {{DOI}: https://doi.org/10.24432/C5XW20}
}

@article{strumbelj2010efficient,
  title={An efficient explanation of individual classifications using game theory},
  author={{\v{S}}trumbelj, Erik and Kononenko, Igor},
  journal={The Journal of Machine Learning Research},
  volume={11},
  pages={1--18},
  year={2010},
  publisher={JMLR. org}
}

@inproceedings{garreau2020explaining,
  title={Explaining the explainer: A first theoretical analysis of LIME},
  author={Garreau, Damien and Luxburg, Ulrike},
  booktitle={International conference on artificial intelligence and statistics},
  pages={1287--1296},
  year={2020},
  organization={PMLR}
}

@book{bishop2006pattern,
  title={Pattern recognition and machine learning},
  author={Bishop, Christopher M.},
  year={2006},
  publisher={Springer}
}

@article{galar2011overview,
  title={An overview of ensemble methods for binary classifiers in multi-class problems: Experimental study on one-vs-one and one-vs-all schemes},
  author={Galar, Mikel and Fern{\'a}ndez, Alberto and Barrenechea, Edurne and Bustince, Humberto and Herrera, Francisco},
  journal={Pattern Recognition},
  volume={44},
  number={8},
  pages={1761--1776},
  year={2011},
  publisher={Elsevier}
}

@article{sokolova2009systematic,
  title={A systematic analysis of performance measures for classification tasks},
  author={Sokolova, Marina and Lapalme, Guy},
  journal={Information processing \& management},
  volume={45},
  number={4},
  pages={427--437},
  year={2009},
  publisher={Elsevier}
}

@article{huang2005using,
  title={Using AUC and accuracy in evaluating learning algorithms},
  author={Huang, Jin and Ling, Charles X},
  journal={IEEE Transactions on knowledge and Data Engineering},
  volume={17},
  number={3},
  pages={299--310},
  year={2005},
  publisher={IEEE}
}

@article{powers2020evaluation,
  title={Evaluation: from precision, recall and F-measure to ROC, informedness, markedness and correlation},
  author={Powers, David MW},
  journal={arXiv preprint arXiv:2010.16061},
  year={2020}
}

@article{brier1950verification,
  title={Verification of forecasts expressed in terms of probability},
  author={Brier, Glenn W},
  journal={Monthly weather review},
  volume={78},
  number={1},
  pages={1--3},
  year={1950}
}

@misc{car_evaluation_19,
  author       = {Bohanec, Marko},
  title        = {{Car Evaluation}},
  year         = {1988},
  howpublished = {{UCI Machine Learning Repository}},
  note         = {{DOI}: https://doi.org/10.24432/C5JP48}
}

@misc{dry_bean_602,
  title        = {{Dry Bean}},
  year         = {2020},
  howpublished = {{UCI Machine Learning Repository}},
  note         = {{DOI}: https://doi.org/10.24432/C50S4B}
}

@misc{nursery_76,
  author       = {Rajkovic, Vladislav},
  title        = {{Nursery}},
  year         = {1989},
  howpublished = {{UCI Machine Learning Repository}},
  note         = {{DOI}: https://doi.org/10.24432/C5P88W}
}

@book{murphy2012machine,
  title={Machine learning: a probabilistic perspective},
  author={Murphy, Kevin P},
  year={2012},
  publisher={MIT press}
}

@article{vaswani2017attention,
  title={Attention is all you need},
  author={Vaswani, A},
  journal={Advances in Neural Information Processing Systems},
  year={2017}
}

@inproceedings{he2019rethinking,
  title={Rethinking imagenet pre-training},
  author={He, Kaiming and Girshick, Ross and Doll{\'a}r, Piotr},
  booktitle={Proceedings of the IEEE/CVF international conference on computer vision},
  pages={4918--4927},
  year={2019}
}

@article{NCI1NCI109,
  title={Comparison of descriptor spaces for chemical compound retrieval and classification},
  author={Wale, Nikil and Watson, Ian A and Karypis, George},
  journal={Knowledge and Information Systems},
  volume={14},
  pages={347--375},
  year={2008},
  publisher={Springer}
}

@article{Protein,
  title={Protein function prediction via graph kernels},
  author={Borgwardt, Karsten M and Ong, Cheng Soon and Sch{\"o}nauer, Stefan and Vishwanathan, SVN and Smola, Alex J and Kriegel, Hans-Peter},
  journal={Bioinformatics},
  volume={21},
  number={suppl\_1},
  pages={i47--i56},
  year={2005},
  publisher={Oxford University Press}
}

@inproceedings{Fey/Lenssen/2019,
  title={Fast Graph Representation Learning with {PyTorch Geometric}},
  author={Fey, Matthias and Lenssen, Jan E.},
  booktitle={ICLR Workshop on Representation Learning on Graphs and Manifolds},
  year={2019},
}

@article{lin2024machine,
  title={Machine learning-assisted dual-atom sites design with interpretable descriptors unifying electrocatalytic reactions},
  author={Lin, Xiaoyun and Du, Xiaowei and Wu, Shican and Zhen, Shiyu and Liu, Wei and Pei, Chunlei and Zhang, Peng and Zhao, Zhi-Jian and Gong, Jinlong},
  journal={Nature Communications},
  volume={15},
  number={1},
  pages={8169},
  year={2024},
  publisher={Nature Publishing Group UK London}
}

@article{rahal2024rise,
  title={The rise of machine learning in the academic social sciences},
  author={Rahal, Charles and Verhagen, Mark and Kirk, David},
  journal={AI \& SOCIETY},
  volume={39},
  number={2},
  pages={799--801},
  year={2024},
  publisher={Springer}
}

@article{jordan2015machine,
  title={Machine learning: Trends, perspectives, and prospects},
  author={Jordan, Michael I and Mitchell, Tom M},
  journal={Science},
  volume={349},
  number={6245},
  pages={255--260},
  year={2015},
  publisher={American Association for the Advancement of Science}
}

@article{carleo2019machine,
  title={Machine learning and the physical sciences},
  author={Carleo, Giuseppe and Cirac, Ignacio and Cranmer, Kyle and Daudet, Laurent and Schuld, Maria and Tishby, Naftali and Vogt-Maranto, Leslie and Zdeborov{\'a}, Lenka},
  journal={Reviews of Modern Physics},
  volume={91},
  number={4},
  pages={045002},
  year={2019},
  publisher={APS}
}

@article{cawley2010over,
  title={On over-fitting in model selection and subsequent selection bias in performance evaluation},
  author={Cawley, Gavin C and Talbot, Nicola LC},
  journal={The Journal of Machine Learning Research},
  volume={11},
  pages={2079--2107},
  year={2010},
  publisher={JMLR. org}
}

@inproceedings{yang2023test,
  title={Test accuracy vs. generalization gap: Model selection in nlp without accessing training or testing data},
  author={Yang, Yaoqing and Theisen, Ryan and Hodgkinson, Liam and Gonzalez, Joseph E and Ramchandran, Kannan and Martin, Charles H and Mahoney, Michael W},
  booktitle={Proceedings of the 29th ACM SIGKDD Conference on Knowledge Discovery and Data Mining},
  pages={3011--3021},
  year={2023}
}

@inproceedings{van1979information,
  title={Information retrieval: theory and practice},
  author={Van Rijsbergen, C},
  booktitle={Proceedings of the joint IBM/University of Newcastle upon tyne seminar on data base systems},
  volume={79},
  pages={1--14},
  year={1979}
}

@article{rumelhart1986learning,
  title={Learning representations by back-propagating errors},
  author={Rumelhart, David E and Hinton, Geoffrey E and Williams, Ronald J},
  journal={nature},
  volume={323},
  number={6088},
  pages={533--536},
  year={1986},
  publisher={Nature Publishing Group UK London}
}

@inproceedings{knerr1990single,
  title={Single-layer learning revisited: a stepwise procedure for building and training a neural network},
  author={Knerr, Stefan and Personnaz, L{\'e}on and Dreyfus, G{\'e}rard},
  booktitle={Neurocomputing: algorithms, architectures and applications},
  pages={41--50},
  year={1990},
  organization={Springer}
}

@book{hosmer2013applied,
  title={Applied logistic regression},
  author={Hosmer Jr, David W and Lemeshow, Stanley and Sturdivant, Rodney X},
  year={2013},
  publisher={John Wiley \& Sons}
}

@article{kipf2016semi,
  title={Semi-supervised classification with graph convolutional networks},
  author={Kipf, Thomas N and Welling, Max},
  journal={arXiv preprint arXiv:1609.02907},
  year={2016}
}

@misc{titanic_openml_40945,
  author = {Harrell, Jr., Frank E. and Cason, Thomas},
  title = {{Titanic Passenger Data}},
  year = {2017},
  howpublished = {OpenML, dataset 40945, version 1},
  url = {https://www.openml.org/d/40945},
  note = {Source documentation: https://hbiostat.org/data/repo/titanic.html}
}

@misc{breast_cancer_wisconsin_diagnostic_17,
  author = {Wolberg, William and Mangasarian, Olvi and Street, Nick and Street, W.},
  title = {{Breast Cancer Wisconsin (Diagnostic)}},
  year = {1993},
  howpublished = {{UCI Machine Learning Repository}},
  doi = {10.24432/C5DW2B}
}

@misc{mnist_openml_554,
  author = {LeCun, Yann and Cortes, Corinna and Burges, Christopher J. C.},
  title = {{MNIST}: Handwritten Digit Images},
  year = {2014},
  howpublished = {OpenML, dataset 554, version 1},
  url = {https://www.openml.org/d/554}
}

@misc{har_smartphones_240,
  author = {Reyes-Ortiz, Jorge and Anguita, Davide and Ghio, Alessandro and Oneto, Luca and Parra, Xavier},
  title = {{Human Activity Recognition Using Smartphones}},
  year = {2013},
  howpublished = {{UCI Machine Learning Repository}},
  doi = {10.24432/C54S4K}
}

@misc{magic_gamma_159,
  author = {Bock, R.},
  title = {{MAGIC Gamma Telescope}},
  year = {2004},
  howpublished = {{UCI Machine Learning Repository}},
  doi = {10.24432/C52C8B}
}

@inproceedings{morris2020tudataset,
  author = {Morris, Christopher and Kriege, Nils M. and Bause, Franka and Kersting, Kristian and Mutzel, Petra and Neumann, Marion},
  title = {{TUDataset}: A Collection of Benchmark Datasets for Learning with Graphs},
  booktitle = {ICML Workshop on Graph Representation Learning and Beyond},
  year = {2020},
  url = {https://arxiv.org/abs/2007.08663}
}

@article{shervashidze2011weisfeiler,
  author = {Shervashidze, Nino and Schweitzer, Pascal and van Leeuwen, Erik Jan and Mehlhorn, Kurt and Borgwardt, Karsten M.},
  title = {{Weisfeiler--Lehman} Graph Kernels},
  journal = {Journal of Machine Learning Research},
  volume = {12},
  number = {77},
  pages = {2539--2561},
  year = {2011},
  url = {https://jmlr.org/papers/v12/shervashidze11a.html}
}

@article{kazius2005mutagenicity,
  author = {Kazius, Jeroen and McGuire, Ross and Bursi, Roberta},
  title = {Derivation and Validation of Toxicophores for Mutagenicity Prediction},
  journal = {Journal of Medicinal Chemistry},
  volume = {48},
  number = {1},
  pages = {312--320},
  year = {2005},
  doi = {10.1021/jm040835a}
}

@inproceedings{riesen2008iam,
  author = {Riesen, Kaspar and Bunke, Horst},
  title = {{IAM} Graph Database Repository for Graph Based Pattern Recognition and Machine Learning},
  booktitle = {Structural, Syntactic, and Statistical Pattern Recognition},
  pages = {287--297},
  year = {2008},
  publisher = {Springer},
  doi = {10.1007/978-3-540-89689-0_33}
}

@misc{imvpy2026,
  key = {imvpy},
  title = {{imvpy}: InterModel Vigorish for probabilistic model comparison and attribution},
  year = {2026},
  howpublished = {Python package, version 1.2.0},
  url = {https://pypi.org/project/imvpy/1.2.0/}
}

@inproceedings{covert2020sage,
  title={Understanding Global Feature Contributions With Additive Importance Measures},
  author={Covert, Ian and Lundberg, Scott M. and Lee, Su-In},
  booktitle={Advances in Neural Information Processing Systems},
  volume={33},
  pages={17212--17223},
  year={2020},
  publisher={Curran Associates, Inc.}
}

\end{document}